\documentclass[twoside,journal]{IEEEtran}
\usepackage{my_layout}
\usepackage{my_macros}

\title{A High Efficiency Superconducting On-chip Filterbank with Directional Filters for Integral Field Units in the Sub-millimeter Regime}

\author{
    Louis H. Marting,
    Kenichi Karatsu,
    Leon G.G. Olde Scholtenhuis,
    Shahab O. Dabironezare, \IEEEmembership{Fellow, IEEE},
    Alejandro Pascual Laguna,
    Arend Moerman,
    David J. Thoen,
    A. J. (Ton) van der Linden,
    Akira Endo,
    and~Jochem J.A. Baselmans
    \thanks{
        Manuscript received xxxxx; revised xxxxx; accepted xxxxx. Date of publication xxxxx; date of current version xxxxx.
        This work was supported by the ESA Contract No 4000139563/22/NL/AS, “Development of a Multi-Chroic, Dual Polarizaiton, Wide-Band On-Chip Array for CMB Spectroscopy”.
        This work was also supported by the European Union (ERC Consolidator Grant No. 101043486 TIFUUN).
        The work of A. Pascual Laguna was supported by the Juan de la Cierva grant JDC2023-051842-I funded by the Spanish MCIN/AEI/10.13039/501100011033.
        (\textit{Corresponding author: Louis H. Marting})
    }
    \thanks{
        Louis H. Marting, Leon G.G. Olde Scholtenhuis, Arend Moerman, Shahab O. Dabironezare, Akira Endo, and Jochem J.A. Baselmans are with the Terahertz Sensing Group, Delft University of Technology, Delft, 2628 CD, The Netherlands.
    }
    \thanks{
        Louis H. Marting, Kenichi Karatsu, Shahab O. Dabironezare, David J. Thoen, A. J. (Ton) van der Linden, and Jochem J.A. Baselmans are with the Space Research Organization Netherlands (SRON), Niels Bohrweg 4, Leiden, 2333 CA, The Netherlands.
    }
    \thanks{
        Alejandro Pascual Laguna is with Centro de Astrobiolog\'ia (CSIC-INTA), Torrej\'on de Ardoz, 28850, Spain.
    }
    \thanks{Jochem J.A. Baselmans is with the Institute of Pysics I, University of Cologne, Z\"ulpicher Stra\ss e 77, Cologne, 50937, Germany.}
    \thanks{
        The reproduction package for this paper is available at
        \url{https://doi.org/10.5281/zenodo.18873636}.
    }
}

\begin{document}

\maketitle

\begin{abstract}
Integrated superconducting spectrometers are developing to the point that they are enabling integral field units, providing large area spectral mapping capabilities for astronomy in the sub-millimeter band.
However, these integral field units are only worthwhile if they have a high efficiency, but to date the efficiency of on-chip filterbanks has been quite poor.
Here we demonstrate a filterbank with high efficiency by using directional filters.
Using a cryogenic thermal load and a noise measurement in combination with a continuous-wave terahertz source to obtain the spectral response of the filters, we are able to accurately measure the filterbank efficiency, accounting for all quasi-optical elements within our setup.
We experimentally obtain an average peak coupling efficiency to the detectors of 75\% in a filterbank that sparsely samples between 125 GHz to 220 GHz using filters with a mean loaded quality factor of 19.6.
Our results demonstrate that a filterbank with a high efficiency is achievable using directional filters, giving a clear route towards efficient integral field units.
\end{abstract}


\section{Introduction}
\IEEEPARstart{T}{he} integrated superconducting spectrometer (ISS) \autocite{taniguchiDESHIMA20Development2022,redfordSuperSpecOnChipSpectrometer2022,robsonSimulationDesignOnChip2022,bensonSpectralCharacterizationPerformance2025,mirzaeiUspecSpectrometersEXCLAIM2020} is maturing to the point that integral field units (IFUs) with ISSs are becoming feasible at (sub-)millimeter wavelengths \autocite{jovanovic2023AstrophotonicsRoadmap2023,dibertOnSkyAtmosphericCalibration2025,bensonSpectralCharacterizationPerformance2025,kohnoLargeFormatImaging2020}.
These IFUs will provide a large field-of-view and a broad spectral bandwidth with moderate spectral resolution, enabling several observing techniques such as (sub-)millimeter-wave line-intensity mapping \autocite{royCrosscorrelationTechniquesMitigate2024, karoumpisCIILineIntensity2024,marcuzzoConstrainingIILuminosity2025,fronenbergForecastsStatisticalInsights2024}, unbiased surveys of dusty line-emitting galaxies \autocite{vankampenAtacamaLargeAperture2024a,kovacsConceptIntegralField2025}, galaxy-cluster analysis using the Sunyaev-Zeldovich effect \autocite{dimascoloAtacamaLargeAperture2025,mroczkowskiAstrophysicsSpatiallySpectrally2019}, and spectro-polarimetry of the CMB \autocite{delabrouilleMicrowaveSpectroPolarimetryMatter2019, sptpolcollaborationDetectionBModePolarization2013}.
Thanks to the integration on a wafer, ISS-based IFUs are far more compact and scalable compared to their quasi-optical analogs (e.g. FTS \autocite{theconcertocollaborationWideFieldofviewLowresolution2020, huCONCERTOAPEXOnsky2024, desertContinuumCOWater2025}, FPI \autocite{cothardDesignCCATprimeEpoch2020}, grating \autocite{ferkinhoffDesignFirstlightPerformance2012,bradfordZSpecBroadbandMillimeterwave2004}), and do not require any opto-mechanics.
Hence, the ISS-based IFUs are regarded as the enabling technology for large-format 3D surveyors on existing telescope facilities (e.g. ASTE, APEX, SPT \autocite{youngDesignPerformanceSPTSLIM2025a}), and on future large telescopes, like CCAT/FYST \autocite{parshleyCCATprimeFredYoung2022} or the proposed 50m AtLAST telescope \autocite{mroczkowskiConceptualDesign50meter2025}.

The IFU is an array of spaxels (spatial pixels), each of which is an independent ISS consisting of a lens-antenna, a superconducting on-chip filterbank with tens to hundreds of spectral channels, and a kinetic inductance detector (KID) at the output of each filter.
The array of KIDs is read out making use of frequency multiplexing, allowing \tild 1000 KIDs per readout line \autocite{vanrantwijkMultiplexedReadout1000Pixel2016}.
The compactness of the filterbanks and the simple multiplexing of KIDs is exactly what enables the scaling of IFUs toward larger arrays.

However, upscaling to a large IFU would only be worthwhile if the individual ISS spaxels have a high sensitivity.
To achieve a high sensitivity, each ISS spaxel must be 1) photon-noise limited, i.e., no instrumental noise in addition to the intrinsic thermal noise of the photons from the foreground atmosphere, and 2) have a high coupling efficiency from the lens to the KIDs.
The ISS technology has proven photon-noise limited operation on a ground-based telescope \autocite{endoFirstLightDemonstration2019}, but the efficiency has been low, limiting the total sensitivity, and therefore the surveying speed, of the system.

Existing on-chip spectrometers lose more than \qty{75}{\percent} of the incoming signal at the peak coupling efficiency of individual channels.
This is due to dielectric losses in the filterbank \autocite{buijtendorpVibrationalModesOrigin2025}, overlapping sidebands from the filters, and due to the fundamental \qty{50}{\percent} coupling limitation for half-wave resonator filters \autocite{martingDirectionalFilterDesign2024a}, which are typically used in on-chip filterbanks \cite{redfordSuperSpecOnChipSpectrometer2022,taniguchiDESHIMA20Development2022,robsonSimulationDesignOnChip2022}. The current highest measured efficiency of a densely sampled on-chip filterbank is an average of \qty{16}{\percent} peak coupling efficiency with a Q-factor of 340 across 334 channels \autocite{karatsuDESHIMA202004002026}, or \qty{27}{\percent} peak efficiency when arrayed in a spectrally sparse configuration \autocite{pascuallagunaTerahertzBandPassFilters2021}.
The maximum coupling efficiency can ideally be improved to \qty{100}{\percent} by using a coherent filter architecture such as a directional filter, instead of a half-wave resonator structure \autocite{martingDirectionalFilterDesign2024a}.

Here we experimentally demonstrate the high coupling efficiency of directional filters.
We fabricated a filterbank with a spectral coverage from 125 GHz to 220 GHz and a loaded quality factor of 25. We place the filters in our filterbank such that they sparsely fill the frequency space, giving a total of eight channels with very limited overlap in the filter bands, allowing us to study the filter behavior in detail without the added complexity of filter overlap.
The design and fabrication of our device are discussed in \cref{sec:design} and \cref{sec:fabrication}, respectively.
The efficiency of the filterbank was measured using a combination of a cryogenic blackbody measurement and a continuous-wave terahertz source.
The methods for our measurement and analysis are described in \cref{sec:methods} and the results are discussed in \cref{sec:results}.

\section{Device Design}\label{sec:design}

\begin{figure*}[!t]
    \centering
    \includegraphics[width=\textwidth]{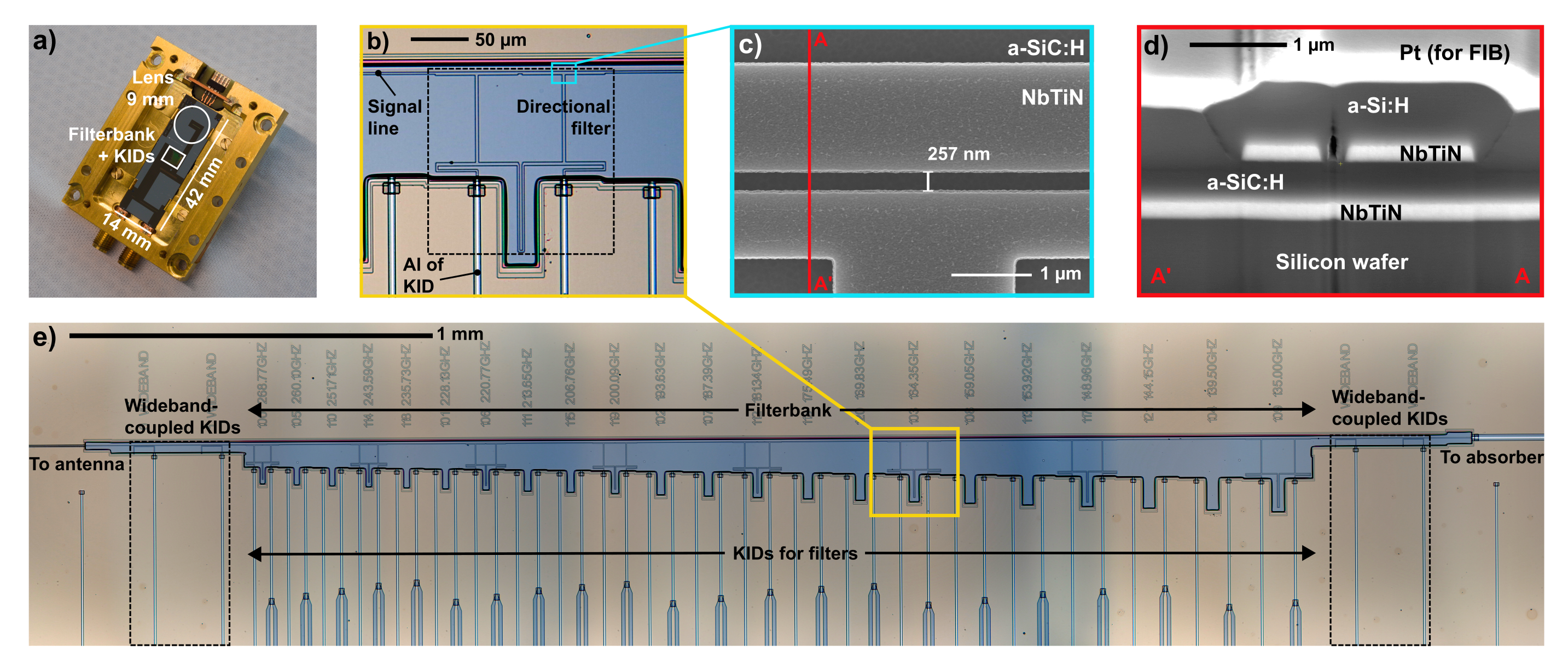}
    \caption{\textbf{a)} Fabricated chip (42x14 mm) in holder. \textbf{b)} Optical micrograph of a single directional filter. See \cref{fig:design_filter} for a schematic of the filter design. \textbf{c)} A scanning electron microscope (SEM) micrograph of part of the coupler. (without top dielectric) \textbf{d)} A SEM micrograph of a focussed ion-beam (FIB) cross-section of the coupler, showing the stratification of the dielectric layers and the metals. \textbf{e)} Stitched optical micrograph of the fabricated sparse filterbank. The highest frequency filter, which is the first in line from the antenna, is to the left. Note the pairs of wideband-coupled KIDs at the start and end of the filterbank.}
    \label{fig:design}
\end{figure*}

\begin{figure}[!t]
    \centering
    \includegraphics[width=\linewidth]{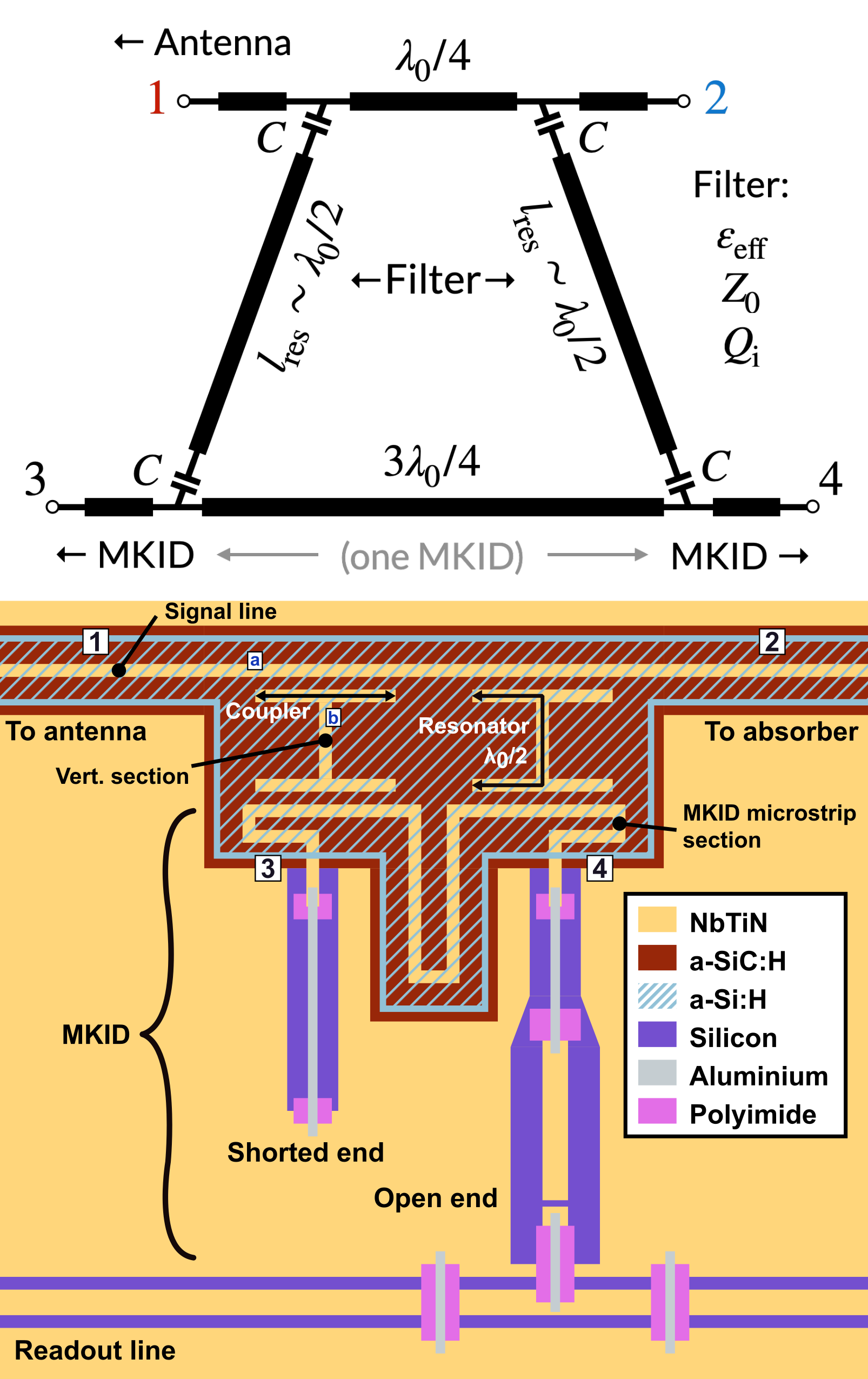}
    \caption{Transmission line circuit of the directional filter (reproduced from \autocite{martingDirectionalFilterDesign2024a}) and its implementation in our device. The KID is connected to port 3 and 4 of the filter. The bottom part of the filter is part of the quarter-wave KID.}
    \label{fig:design_filter}
\end{figure}

\begin{figure}[h]
    \centering
    \includegraphics[width=\linewidth]{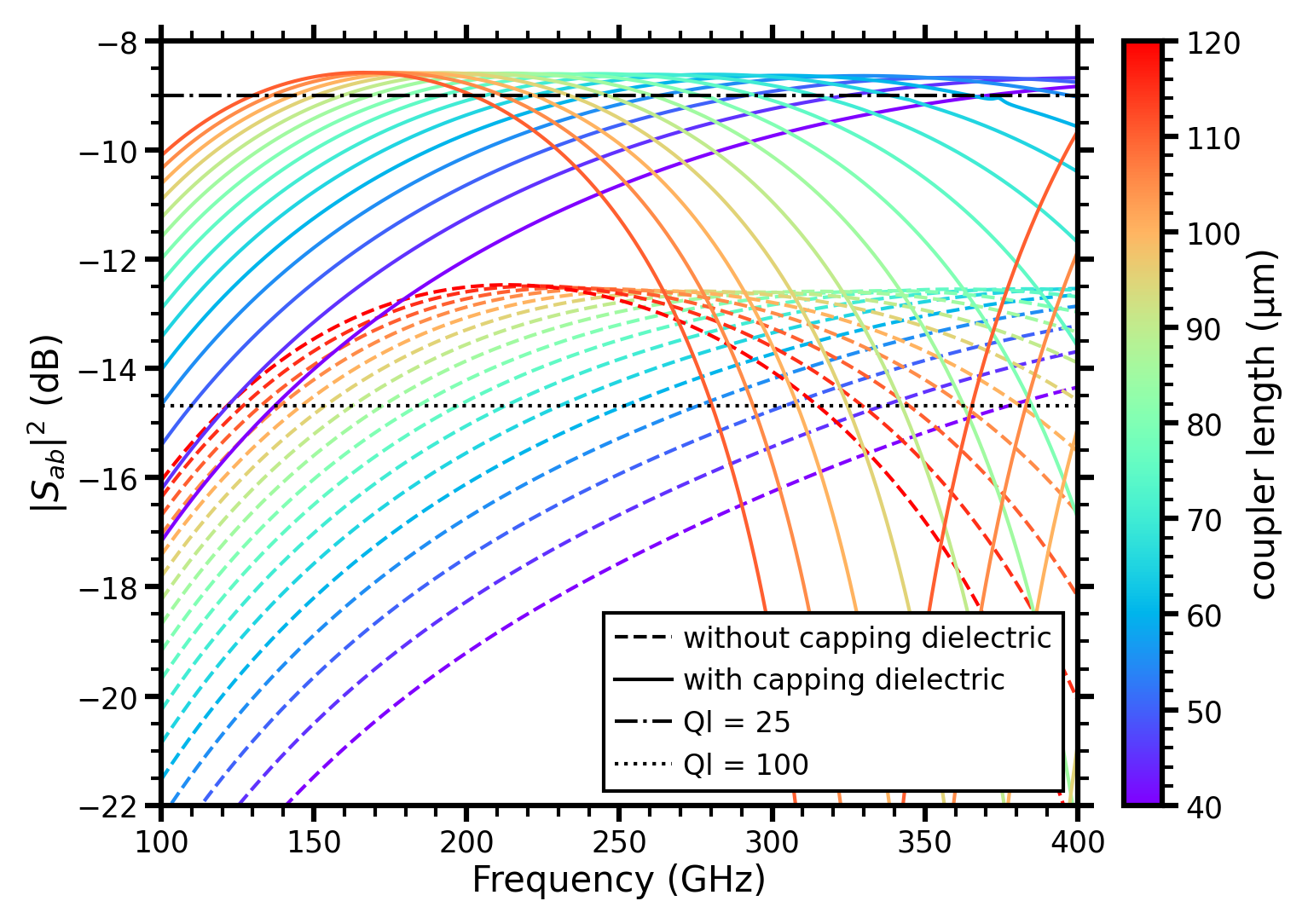}
    \caption{A SONNET simulation of a co-planar microstrip coupler with and without capping dielectric. The horizontal dash-dot and dotted lines indicate the coupling strength required for a certain $\Ql$ following \cref{eq:SQc}. The coupling structure, with the definitions of ports a and b, can be found in \cref{fig:design_filter}.}
    \label{fig:coupler}
\end{figure}

The device is shown in \cref{fig:design}a. The chip and holder are based very closely on the DESHIMA 1.0 design \autocite{endoFirstLightDemonstration2019,endoWidebandOnchipTerahertz2019} (twinslot antenna, lens and holder), with a sparse filterbank using a microstrip based filter design using fabrication technology similar to DESHIMA 2.0 \autocite{taniguchiDESHIMA20Development2022,karatsuDESHIMA202004002026}.
In this work we realize a sparse configuration of a filterbank that is designed to cover a bandwidth of one octave from 135 GHz to 270 GHz with a filter spacing $\Delta f / f$ equal to a spectral resolution R of 30 and a loaded quality factor $\Ql$ of 25, which makes the filterbank design slightly oversampled.
The number of filters for this spectral resolution is given by
\begin{equation}
	N = \left\lceil \frac{\ln(f_\mathrm{max}/f_\mathrm{min})}{\ln(1 + R^{-1})} \right\rceil,
\end{equation}
where $\lceil \cdot \rceil$ is the ceiling function, and results in 22 filters in the band.
Each resonator has a resonance given by
\begin{equation}
    f_n = f_\mathrm{min}(1 + R^{-1})^{n-1} , \text{where } n \in [1,\dots,N].
\end{equation}

We use a sparse filterbank configuration to be able to study the individual filter responses without being affected by the overlapping sidebands of neighboring filters \autocite{martingDirectionalFilterDesign2024a}. This is realized by only patterning one in every three filters, which can be seen in the microscope image in \cref{fig:design}e.

\begin{figure*}[!t]
    \centering
    \includegraphics[width=\textwidth]{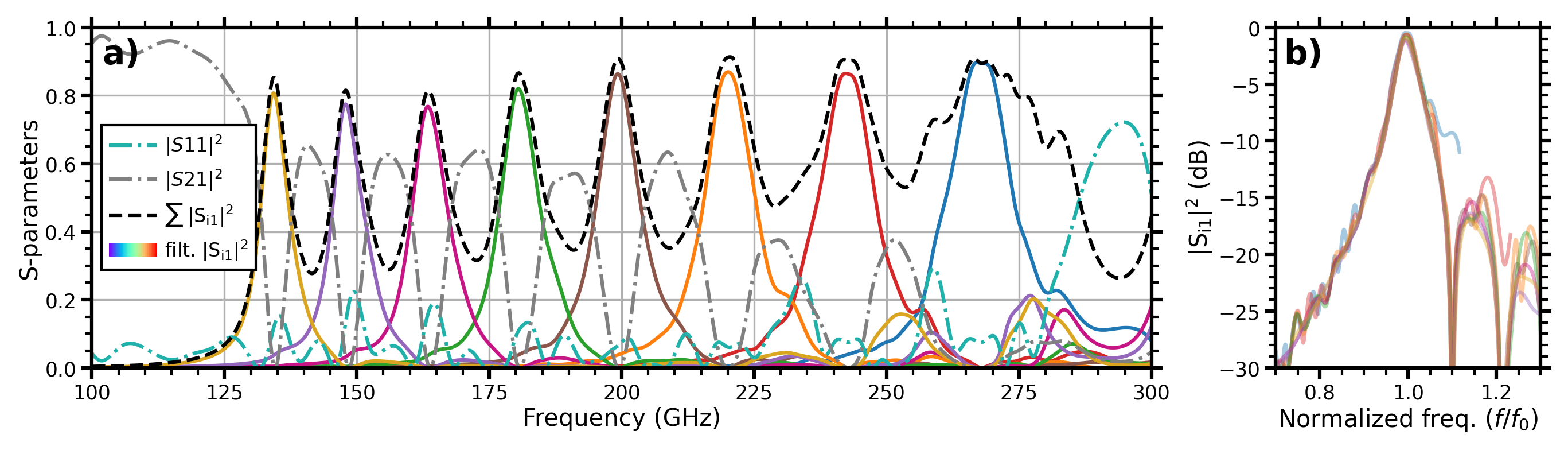}
    \caption{\textbf{a)} Simulation of the sparse filterbank design using a transmission line model \autocite{martingDirectionalFilterDesign2024a,pascuallagunaTerahertzBandPassFilters2021}. \textbf{b)} Stacked filter responses with normalized frequency. The dips in the lorentzian-like response at fixed intervals are due to filter suckouts of preceding filters.}
    \label{fig:simulated_fb}
\end{figure*}

\subsection{Filter design}
The filterbank consists of directional filters \autocite{martingDirectionalFilterDesign2024a}.
A circuit model and a schematic of the directional filter as implemented in our device is given in \cref{fig:design_filter}.
The directional filter creates an impedance matched condition on- and off-resonance by using a phase-coherent structure.
This leads to a more efficient coupling to the KID, overcoming the \qty{50}{\percent} maximum peak coupling efficiency of half-wave filters.

The filters are designed with a loaded quality factor $\Ql$ of 25 to minimize the influence of dielectric losses on our experiment \autocite{pascuallagunaTerahertzBandPassFilters2021,buijtendorpVibrationalModesOrigin2025}.
The coupling strength $|S_\mathrm{ab}|^2$ across a single coupling element and the coupling quality factor $\Qc$ are related with
\begin{equation}\label{eq:SQc}
	|S_\mathrm{ab}|^2 = \frac{2 \pi}{\Qc},
\end{equation}
and the loaded quality factor of a resonator surrounded by two of these couplers is given by
\begin{equation}\label{eq:QlQcQi}
	\frac{1}{\Ql} = \frac{2}{\Qc} + \frac{1}{\Qi},
\end{equation}
where $\Qi$ represents the internal losses of the resonator. For our dielectric materials, detailed in \cref{sec:fabrication}, we expect a $\Qi$ between \numrange{1e3}{1e4}.
Since $\Qc \ll \Qi$, the internal losses can be neglected and, by combining \cref{eq:SQc} and \cref{eq:QlQcQi}, it can be found that for a loaded quality factor $\Ql$ of 25 a coupling strength $|S_\mathrm{ab}|^2 = \qty{-9}{\dB}$ is required.

To reach this required coupling strength, we adapt the DESHIMA 2.0 design \autocite{karatsuDESHIMA202004002026} by adding a top dielectric on the microstrip co-planar coupler.
A schematic drawing of the coupler in the filter is shown in \cref{fig:design_filter} with micrographs of the realized coupler in \cref{fig:design}c-d.
We find that we need a gap of \qty{250}{\nano\meter} and an a-Si capping layer of \qty{800}{\nano\meter} to reach  $|S_\mathrm{ab}|^2 = \qty{-9}{\dB}$ using SONNET simulations, as shown in \cref{fig:coupler}.

Furthermore, this simulation is used to find the exact coupling bar length to reach the desired coupling strength at each individual filter resonance frequency.
Note that the resonators in the filterbank have an electrical length of $\lambda/2$, where $\lambda$ is the wavelength.
This constraint requires that the coupling bar length should be less than $\lambda/4$.
Since we are near the peak coupling strength of this coupler design, it is only slightly shorter than $\lambda/4$, indicating that we are at the lower limit of $\Ql$ for our current coupling geometry.

Using the material parameters of the NbTiN, a-SiC:H, and a-Si:H (see \cref{sec:fabrication}), we define the filter geometry and electrical properties in \cref{tab:filter_geom}.

\begin{table}
    \centering
    \caption{Geometry and electrical properties of filter elements. See \cref{fig:design_filter} for the definition of the filter elements.}
    \label{tab:filter_geom}
    \begin{tabular}{@{} l r r r @{}}
    \toprule 
        {} & {width (\unit{\micro\meter})} & {$Z_0$ (\unit{\Omega})} & {$\varepsilon_\mathrm{eff}$} \\
    \midrule 
        Signal line & \num{1.3} & \num{60.4} & \num{36.9} \\
        Coupler & \num{0.8} & \num{83.3} & \num{42.1} \\
        Vertical section & \num{2.1} & \num{43.0} & \num{33.8} \\
        KID microstrip section & \num{1.3} & \num{60.4} & \num{36.9} \\
    \bottomrule 
    \end{tabular}
\end{table}

We implemented this filterbank into a transmission line model to predict the frequency response of our device  \autocite{martingDirectionalFilterDesign2024a,pascuallagunaTerahertzBandPassFilters2021}. We used the measured layer thicknesses and widths that were obtained using the micrographs in \cref{fig:design}, and we used the measured material parameters specified in \cref{sec:fabrication}.
The sparse, low spectral resolution filterbank configuration results in an average simulated peak efficiency of \qty{80}{\percent} as shown in \cref{fig:simulated_fb}.
This is a \qty{20}{\percent} deviation from a \qty{100}{\percent} percent peak coupling efficiency for a perfect system \autocite{martingDirectionalFilterDesign2024a}.
This is because at strong coupling, the coupler cannot be approximated by a perfect admittance inverter \autocite{pozarMicrowaveEngineering2012}, causing a small reflection on the signal line at resonance.
Decreasing the admittance of the resonator mitigates this problem slightly, but was not implemented for this device as this would have made the width of the coupling elements too narrow for fabrication.
Furthermore, the residual overlapping sidebands of the filters slightly reduces the peak coupling efficiency due to leakage to these preceding filters.

\subsection{KID design}
The quarter-wave KIDs in our filterbank are illustrated in \cref{fig:design_filter} and have a unique geometry because they also form the impedance match with the directional filter at sub-millimeter frequencies, which is required for its proper operation.
The lossy aluminium-NbTiN (hybrid) CPW lines absorb the sub-millimeter signal and form an input impedance approximately equal to the characteristic impedance of the hybrid lines, as long as the total attenuation is significant \autocite{pozarMicrowaveEngineering2012}, i.e. the reflectance amplitude $\Gamma$ as seen at the ports is sufficiently small.

We have a hybrid CPW line with a complex propagation constant $\gamma = \alpha + j \beta = k_0\sqrt{\varepsilon_\mathrm{eff}} = k_0(\sqrt{\complexnum[output-complex-root = j]{7.41-j1.12}})$ which gives an attenuation of \qty{14}{\dB\per\mm} at \qty{190}{\GHz}.
Its length is \tild\qty{1}{\mm} and is divided between port 3 and 4.
We favour a longer length at port 3, since almost all power from the antenna couples to this port, and we ensure there is sufficient length left at port 4 to reasonably approximate this input inpedance with its characteristic impedance.
Furthermore, we couple port 3 of the filter to the shorted end of the KID to maximize the response by absorbing the signal at its current maximum.
It should also be noted that the NbTiN microstrip section that is part of the filter separates the quasiparticle systems of the two sections of aluminium.
As a consequence of this, the volume of the aluminium at the shorted end is the effective aluminium volume of the KID.
We find that a 2/3 to 1/3 split for ports 3 and 4, respectively, are optimum. This gives an input reflectance amplitude $|\Gamma| \propto e^{-2\alpha l}$ at ports 3 and 4 that is attenuated with \qty{9.5}{\dB} and \qty{4.7}{\dB}, respectively.

Finally, we give the design parameters of the KIDs.
There are eight filter KIDs designed between \qty{5.4}{\GHz} and \qty{6.6}{\GHz} with a $\Qc$ equal to \num{2e4}.
There are four wideband KIDs designed around \qty{6.75}{\GHz} with \qty{30}{\mega\Hz} spacing and a $\Qc$ equal to \num{2e4}.
The sub-millimeter signal is weakly wideband-coupled to two KIDs before the filterbank and two after the filterbank, which we will call `widebands before' and `widebands after,' respectively.
Lastly, there are two blind KIDs at \qty{5.25}{\GHz} and \qty{5.3}{\GHz} with a $\Qc$ equal to \num{5e4}.
One is near the start of the filterbank and one is near the end of the filterbank.

\section{Fabrication}\label{sec:fabrication}
The fabrication route of this chip is nearly identical to the one described in \autocite{hahnleSuperconductingMicrostripLosses2021}, for the filter fabrication, we closely mimic the process described in \autocite{thoenCombinedUltravioletElectronbeam2022}.
The only difference is the addition of an \qty{800}{\nano\meter} a-Si:H layer on top of the filterbank deposited using PECVD at \qty{250}{\degreeCelsius} and patterned using optical contact lithography and etched using an SF6 and O2 plasma in a reactive-ion etcher (RIE).

With the extra layer, the bottom-to-top stratification of the filterbank structure is: 200 nm NbTiN ($T_\mathrm{c} = 14.65 \:\mathrm{K}$, $R_\mathrm{s} = 10.69 \:\Omega/\square$, $\rho = 216 \:\upmu\Omega \text{cm}$) \autocite{thoenSuperconductingNbTinThin2017}, 500 nm a-SiC:H ($\varepsilon_\mathrm{r} = 7.8$, $\tan \delta \approx \num{1e-4}$) \autocite{buijtendorpVibrationalModesOrigin2025}, 200 nm NbTiN, and, 800 nm a-Si:H ($\varepsilon_\mathrm{r} = 10.1$, $\tan \delta \approx \num{8e-4}$) \autocite{buijtendorpCharacterizationLowlossHydrogenated2022,pascuallagunaTerahertzBandPassFilters2021}.
This stratification is also visible in the micrograph in \cref{fig:design}d.
Note that we have two different dielectrics in the stratification, this is solely due to machine availability at the time of fabricating.

These four layers in the filterbank stratification are the first layers that are deposited, patterned and etched on the chip.
The fabrication is completed with \qty{1}{\micro\meter} thick Polyimide patches for the readout line bridges; 40 nm aluminium ($T_\mathrm{c} = 1.26 \:\mathrm{K}$, $R_\mathrm{s} = 0.225 \:\Omega/\square$) for the KIDs and the bridges; and, 40 nm of $\beta$-phase tantalum ($T_\mathrm{c} = 0.65 \:\mathrm{K}$, $R_\mathrm{s} \approx 61 \:\Omega/\square$) on the backside for stray light control \autocite{yatesSurfaceWaveControl2017}.
After fabrication, the chip is diced and the lens with a spacer wafer is glued to the backside of the wafer. The chip is finally mounted in the holder as shown in \cref{fig:design}a.

\section{Methods}\label{sec:methods}
\begin{figure}[h]
    \centering
    \includegraphics[width=\linewidth]{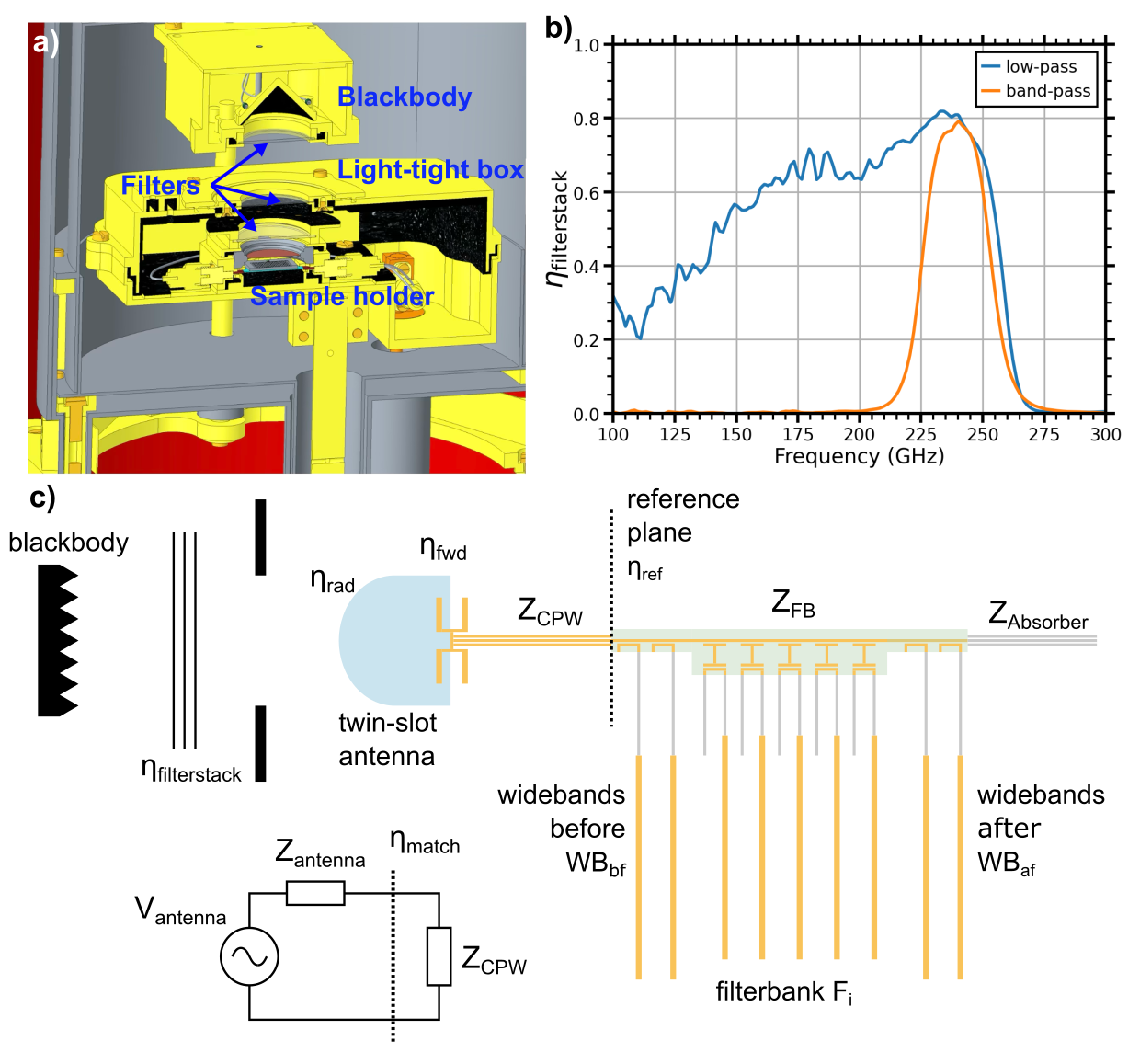}
    \caption{\textbf{a)} Cryogenic blackbody setup with filterstack. \textbf{b)} Total transmission through the filterstack for two different configurations. \textbf{c)} Schematic of the measurement setup, indicating all the parts considered and their efficiencies. See the text for details.}
    \label{fig:experiment}
\end{figure}

To measure the filterbank efficiency, the chip is placed into an adiabatic demagnetization refrigeration (ADR) cryostat cooled to 120 mK.
Our setup is shown in \cref{fig:experiment}a.
This setup is similar to the experiment described in \autocite{devisserFluctuationsElectronSystem2014} and allows us to illuminate the chip and lens with a cryogenic blackbody, capable of radiating with a temperature from 4 K to 40 K.
The efficiency is found using a two step process.

Firstly, we experimentally obtain the coupling efficiency between the detectors and a well defined cryogenic thermal radiator using the photon noise as absolute calibration \autocite{ferrariAntennaCoupledMKID2018}.
This is discussed in \cref{subsec:efficiency}.

Secondly, we measure the frequency response of the filters using a continuous-wave (CW) terahertz source with uncalibrated power output, similar to \autocite{endoFirstLightDemonstration2019}.
This provides the relative frequency response of the filters which is normalized and is applied as the final element in the optical path.
This is discussed in \cref{subsec:normalization}.

\subsection{Optical efficiency}\label{subsec:efficiency}
First, we define the frequency response of the filters from the reference plane in front of the filterbank as shown in \cref{fig:experiment}c, to the KID detector as follows
\begin{equation}\label{eq:eta_i}
    \eta_i(f) = \eta_i^{*} \tilde{\eta}_i(f), \quad \text{where } \max \tilde{\eta}_i(f) = 1.
\end{equation}

In this equation, $\eta_i^{*}$ is the frequency-independent unknown filter efficiency of filter $i$ which we aim to find and $\tilde{\eta}_i(f)$ is the normalized frequency response which we will obtain from our CW terahertz source measurement, which is explained in \cref{subsec:normalization}.

The power absorbed in the aluminium of the KID behind filter $i$ is given by
\begin{equation}\label{eq:Pinc}
	P_\mathrm{inc} = \eta_i^* \int \frac{1}{2}\lambda_0\!{}^{2} \mathrm{B}(f,T_\mathrm{bb})  \eta_\mathrm{ref}(f) \tilde{\eta}_i(f) \;df = \eta_i^* \int S_\mathrm{P}(f)\;df
\end{equation}
Where $\mathrm{B}(f,T_\mathrm{bb})$ is the blackbody specific intensity at a known temperature $T_\mathrm{bb}$ which is converted to a single polarization using the factor 1/2, $\lambda_0$ is the is the free-space wavelength, and $\eta_\mathrm{ref}(f)$ is the transfer efficiency from the blackbody to the reference plane.
We use $S_\mathrm{P}(f)$ in \cref{eq:Pinc} to denote the power spectral density that can be absorbed in the KID for a filter efficiency $\tilde{\eta}_i(f)$ with a peak value of unity.

The transfer efficiency $\eta_\mathrm{ref}(f)$ is decomposed as follows
\begin{equation}\label{eq:eta_ref}
	\eta_\mathrm{ref} = \eta_\mathrm{filterstack} \cdot \eta_\mathrm{rad} \cdot \eta_\mathrm{fwd} \cdot \eta_\mathrm{match},
\end{equation}
where the comprising terms are defined as:
\begin{itemize}
	\item[--] The filterstack efficiency $\eta_\mathrm{filterstack}$ is the transmission of the filterstack used to limit the blackbody radiation to our band of interest.
	It has two configurations which are shown in \cref{fig:experiment}b.
	\item[--] The radiation efficiency $\eta_\mathrm{rad}$ is defined as the fraction of the power radiated over the power used to excite the antenna, accounted for the power reflected at the Si-vacuum lens interface. This is calculated using a quasi-optical simulation \autocite{zhangFourierOpticsTool2021} for our twin-slot antenna with a hyper-hemispherical lens.
	\item[--] The forward efficiency $\eta_\mathrm{fwd}$ defines what fraction of the total radiated power by the antenna is actually radiated in the upper plane of the antenna, in the direction of the lens and blackbody source. This is found from the simulated radiation patterns of the lens-antenna.
	\item[--] The matching efficiency $\eta_\mathrm{match}$ defines how well the antenna impedance is matched to the line impedance of the on-chip CPW line, and is calculated with the Thevenin equivalent circuit shown in \cref{fig:experiment}c.
	Since a twin-slot antenna is resonant, it will have a designed peak matching efficiency at the center of the operation band, which for our antenna is designed to be at 220 GHz.
\end{itemize}

Note that the antenna observes the blackbody through an aperture, which would normally require an spillover efficiency. However, the lens-antenna beam is much narrower than the aperture diameter and $> 99\%$ of the beam sees the blackbody source, therefore the spillover efficiency is neglected.

\begin{figure}[!t]
    \centering
    \includegraphics[width=\linewidth]{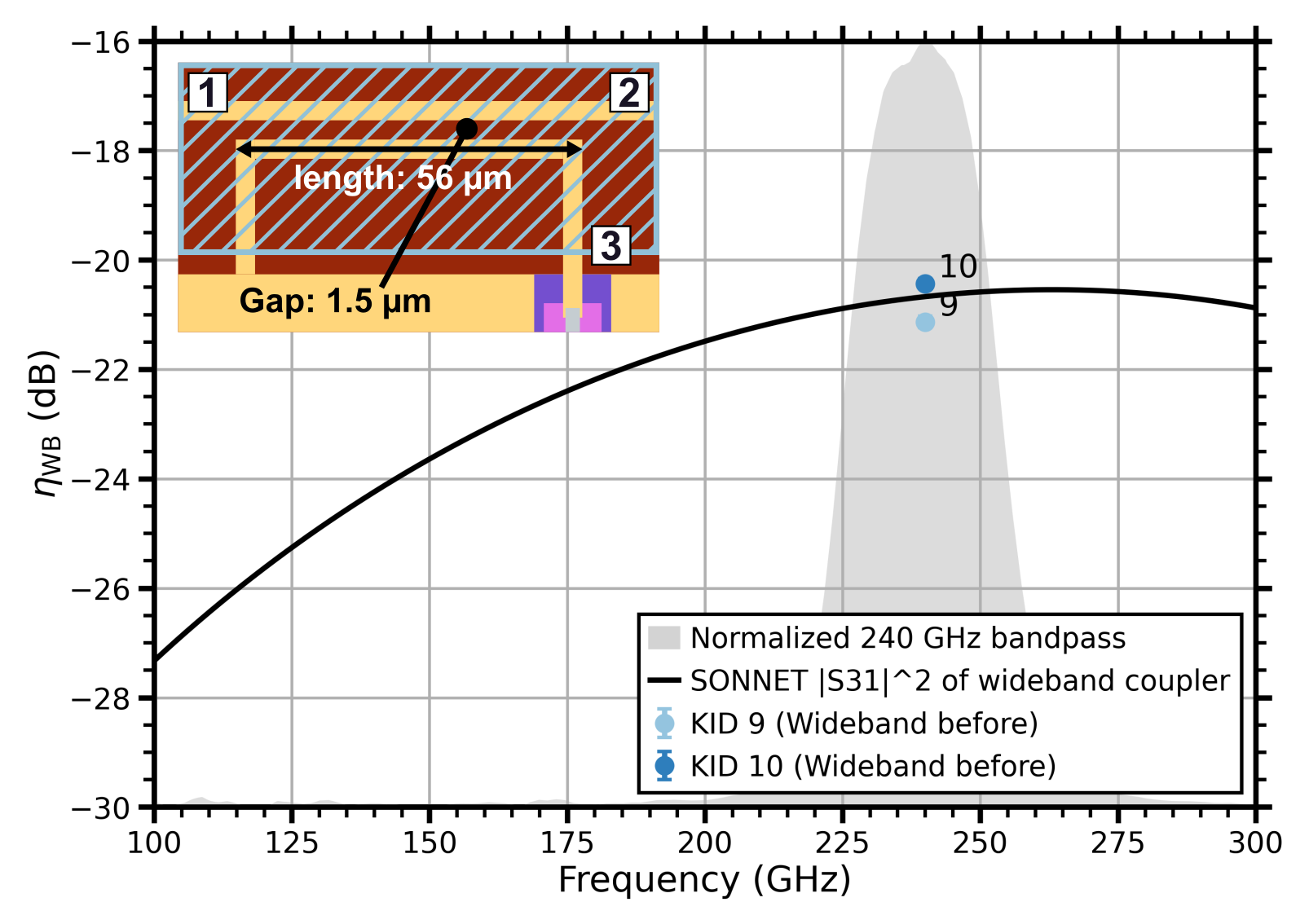}
    \caption{Efficiency of the wideband-coupled KIDs $\eta_\mathrm{WB}$ using the \qty{240}{\GHz} band-pass setup as shown in \cref{fig:experiment}. This is used to verify the calculations of $\eta_\mathrm{ref}$ as described in \cref{subsec:efficiency}. The inset shows the wideband coupler in detail with the port definitions, which is simulated in SONNET. The layers are equal to the legend in \cref{fig:design_filter}.}
    \label{fig:wb_efficiency}
\end{figure}

We verified our calculations of the coupling efficiency to the reference plane $\eta_\mathrm{ref}$ by measuring the efficiency of the wideband KIDs in front of the filterbank using the narrow-band filterstack configuration as shown in \cref{fig:experiment}b.
The results are shown in \cref{fig:wb_efficiency}.
Across the narrow-band filterstack, the coupling strength of the wideband coupler can be considered constant, which means we do not need to make any assumptions for the frequency dependence of the wideband couplers.
Instead, we can directly compare the measured efficiency to the expected efficiency of the wideband-coupled KIDs.
The expected efficiency is obtained from a SONNET simulation using the measured geometry and material parameters.
We explicitly take into account an overetch of \qty{100}{\nano\meter} into the a-SiC:H, wich is filled by a-Si:H, and the thickness of the metals.
The results show that we measure the expected coupling efficiency using our method, validating our calculations for $\eta_\mathrm{ref}$.\\

Since we know $\eta_\mathrm{ref}$ we can experimentally obtain $\eta_i^*$, using the method described by Ferrari et al. \autocite{ferrariAntennaCoupledMKID2018}. This method equates the experimentally measured noise equivalent power ($\mathrm{NEP}_\mathrm{exp}$), which is obtained via a noise measurement and a responsivity measurement, with the theoretical NEP given by the incident power $P_\mathrm{inc}$ as defined in \cref{eq:Pinc}.
The experimental NEP is
\begin{equation}\label{eq:NEPexp}
	\mathrm{NEP}_\mathrm{exp}\!{}^{2}(P_\mathrm{inc}) = S_\theta\left(\frac{d\theta}{dP_\mathrm{inc}}\right)^{-2},
\end{equation}
where $S_\theta$ is the phase noise at white photon noise level (i.e. at sampling frequencies below the quasiparticle lifetime rolloff).
We find the white photon noise level of each KID by fitting the noise using a composite model to isolate it from the 1/f noise and readout noise, the details of our fitting procedure are described in the Appendix.
The responsivity $d\theta / dP_\mathrm{inc}$ is the phase response due to a change in the incident power, which is measured by performing a sweep of the blackbody temperature of a few kelvin around $T_\mathrm{bb}$.

Secondly, the theoretical photon noise limited NEP for a KID calculated from $P_\mathrm{inc}$ in \cref{eq:Pinc} is expressed as

\begin{equation}\label{eq:NEPph}
	\mathrm{NEP}_\mathrm{ph}\!{}^{2} = 2 \int \eta_i^* \; S_\mathrm{P}(f) \left( hf + \frac{2\Delta_\mathrm{Al}}{\eta_\mathrm{pb}}\right) + \eta_i^*{}^2 \; S^{{}^2}_\mathrm{P}(f)\:df.
\end{equation}

Equating \cref{eq:NEPexp} with \cref{eq:NEPph}, and freeing $\eta_i^*$ gives
\begin{equation}\label{eq:eta_star}
    \eta_i^* = \frac{2 \int S_\mathrm{P}(f) \left( hf + \frac{2\Delta_\mathrm{Al}}{\eta_\mathrm{pb}}\right) \;df}{S_\theta \left( d\theta/{\int S_\mathrm{P}(f) \;df} \right)^{-2} - 2 \int S^{{}^2}_\mathrm{P}(f) \;df}
\end{equation}
which is experimentally attainable.

\begin{figure*}[!t]
    \centering
    \includegraphics[width=\linewidth]{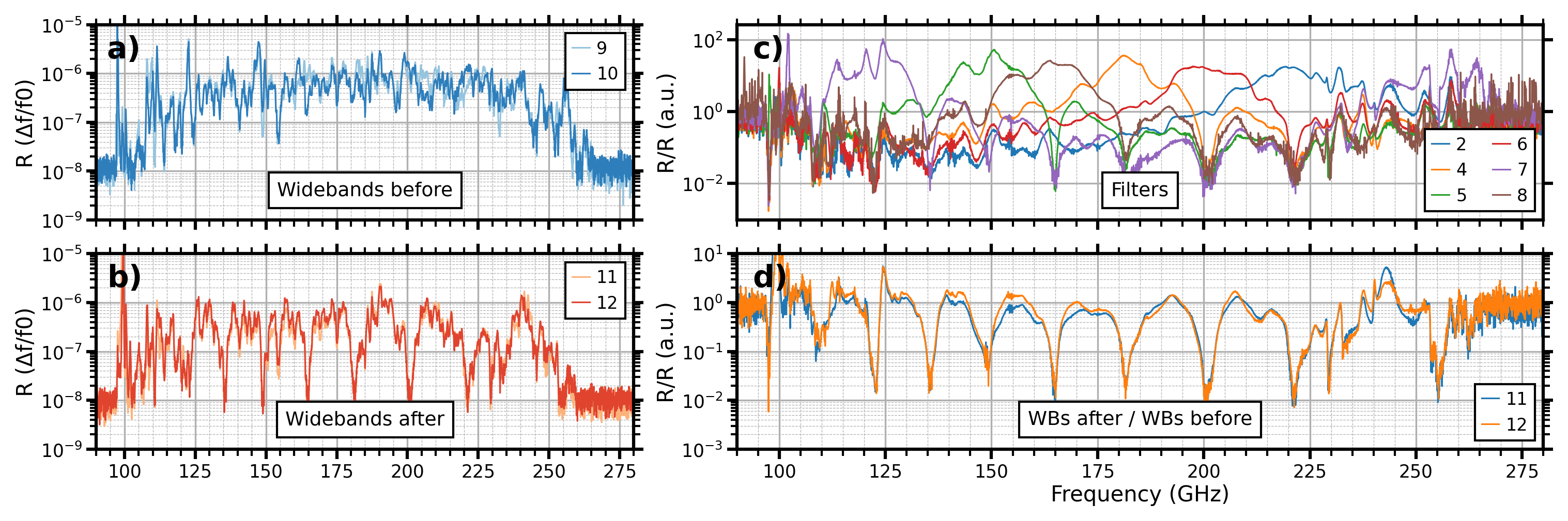}
    \caption{\textbf{a)} Raw measurement data of the wideband couplers before the filterbank, expressed as the observed fractional frequency shift. \textbf{b)} Raw measurement data of the widebands after the filterbank. \textbf{c)} Filter responses normalized to the average response of the widebands before the filterbank. \textbf{d)} Response of the widebands after the filterbank normalized to the average response of the widebands before the filterbank.}
    \label{fig:raw_response}
\end{figure*}

\subsection{Normalization of the filter frequency response}\label{subsec:normalization}
Using a CW terahertz source, we obtain the unnormalized relative frequency response $\tilde{\eta}_i(f)$ as follows.
The CW terahertz source illuminates the chip mounted in a second setup (see \autocite{endoWidebandOnchipTerahertz2019} for details) and the source is swept from \qtyrange{90}{300}{\giga\hertz} with steps of \qty{50}{\mega\hertz}.
Each KID observes the phase response to this signal.
There are a few wideband-coupled KIDs that are weakly coupled to the signal line from the antenna, which can be seen in \cref{fig:experiment}.
These act as a power reference for the on-chip components.
They are used to remove the power fluctuations of the CW terahertz source and other coherent external influences by dividing the responses of the filter KIDs with the average of the wideband-coupled KIDs before the filterbank.
This results in the unnormalized relative frequency response $\tilde{\eta}_i(f)$ at the reference plane.
Furthermore, an estimate of the relative power transmitted across the filterbank can be calculated by dividing the response of the widebands after the filterbank by the response of the widebands before the filterbank.

To get a physically correct efficiency result and complete the optical efficiency calculation, this relative frequency response needs to be normalized to obtain $\max \tilde{\eta}_i(f) = 1$ as defined in \cref{eq:eta_i}.
Ideally, the normalization is simply applied by normalizing each filter with its maximum value, which satisfies the aforementioned condition.
However, as we will observe in the measurements, a standing wave is present on the signal line of the filterbank.
This distorts the filter response with a ripple, making the standard normalization unreliable.

Instead, we use a normalization based on the total area under each filter in combination with its known spectral shape in (semi-)isolation, to integrate out the distortions due to standing waves.
Each filter in (semi-)isolation has a lorentzian filter shape.
We define a standard lorentzian with a loaded quality factor $\Ql$, resonance frequency $f_0$ and a peak intensity $I$ as
\begin{equation}
	\mathcal{L}(f) = I\frac{1}{1 + \left(2\Ql\frac{f-f_0}{f_0}\right)^2}.
\end{equation}
The total area of the lorentzian is
\begin{equation}
	\int_{-\infty}^{\infty} \mathcal{L}(f)\, df = I\frac{\pi f_0}{2  \Ql}
\end{equation}
and the integral across the full-width half maximum (FWHM) is exactly half the total area
\begin{equation}\label{eq:lorentzian_bw}
	\int_\mathrm{FWHM} \mathcal{L}(f)\, df = \frac{1}{2}\int_{-\infty}^{\infty} \mathcal{L}(f)\, df.
\end{equation}
The loaded Q-factor $\Ql$ is calculated using the bandwidth $\Delta f$ at the FWHM.

Besides the lorentzian spectral shape assumption, we assume that the couplers have the same $\Ql$ across the band, as designed. This implies that the area under the frequency-normalized response of all filters is equal. However, we must not neglect a possible systematic difference in $\Ql$ due to the fabrication, making $\Ql$ a constant, but unknown parameter of the lorentzian filter shape. We must therefore first determine $\Ql$ before applying the normalization.

The normalization procedure is as follows.
First, the data of each filter is expressed in terms of its normalized frequency $f/f_0$.
Then, each filter is normalized by its total area.
This normalizes each filter with the same but unknown scaling factor $2 \Ql / \pi$, where the factor $2/\pi$ comes from the lorentzian integral.
To find $\Ql$, the average of all the normalized filters is taken.
From the average normalized response, its FWHM is found by making use of \cref{eq:lorentzian_bw}, finding for what bandwidth the area around the resonance is half the total area.
The $\Ql$ is calculated from the FWHM.
Finally, each individual normalized filter response is scaled with a factor $1/(2 \Ql / \pi)$ to get the true normalized filter response, defined as $\tilde{\eta}_i(f)$ in \cref{eq:eta_i}.
This gives us enough information for the efficiency calculations.

\section{Results \& Discussion}\label{sec:results}

\begin{figure*}[!t]
    \centering
    \includegraphics[width=\linewidth]{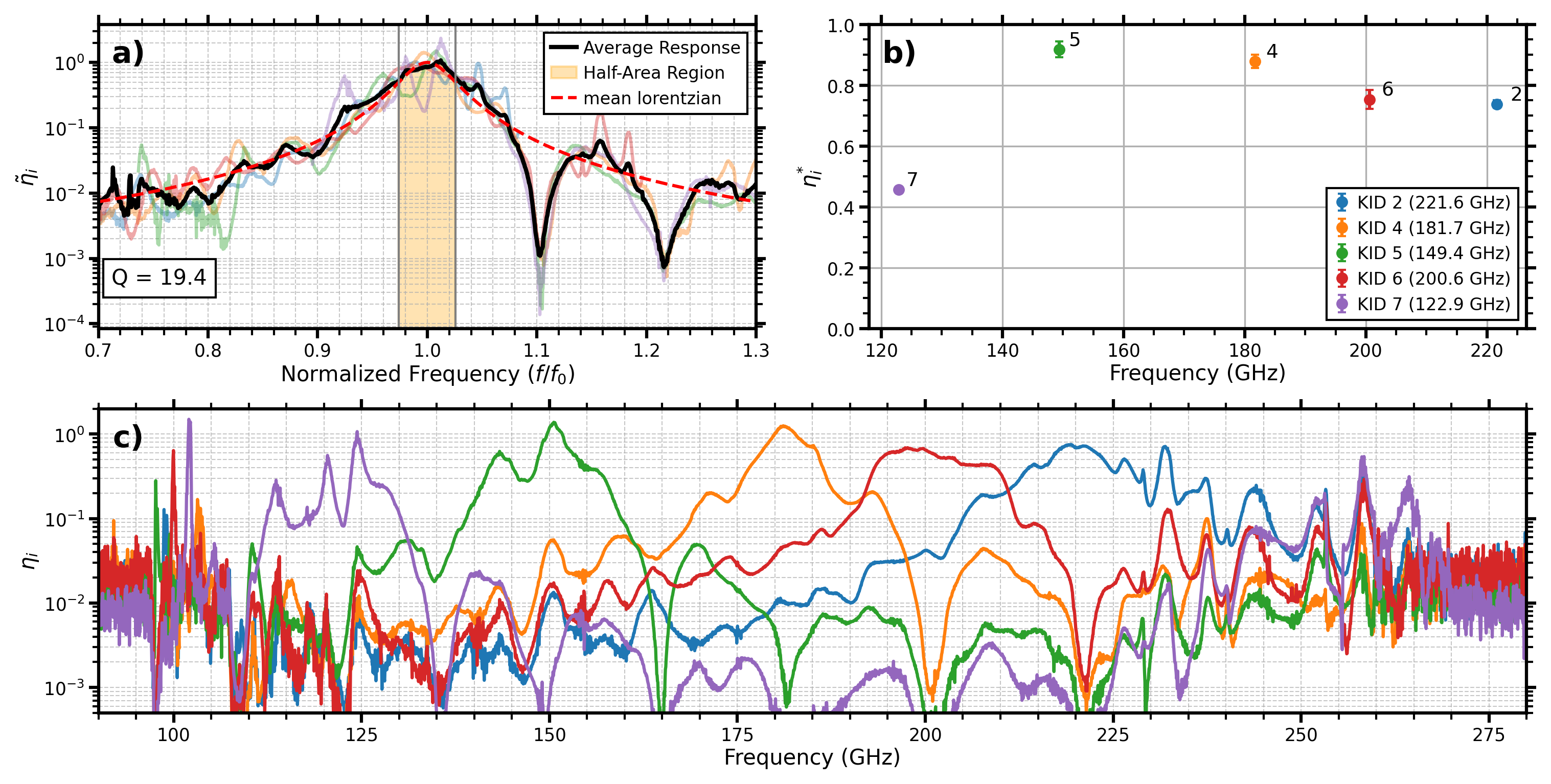}
    \caption{\textbf{a)} Normalized response of each filter using the procedure described in \cref{subsec:normalization}. \textbf{b)} Result of our efficiency calculation as described in \cref{subsec:efficiency} \textbf{c)} Spectral response of each filter across the full filter bandwidth according to the definition in \cref{eq:eta_i} and using the results of subfigures \textbf{a} and \textbf{b}.}
    \label{fig:efficiency}
\end{figure*}

The chip was mounted in its holder and measured.
We found 12 out of 14 KIDs between 5 GHz to 7GHz, with 6 out of 8 filter KIDs, all four wideband KIDs and both blind KIDs found.
We find an average KID $\Qc$ of \num{1.5e4} and an average $\Qi$ of \num{2.7e5}.
We measured the noise of the KIDs in the setup of \cref{fig:experiment} with the blackbody around a mean temperature of $T_\mathrm{bb} = \qty{36}{\kelvin}$ and we performed a blackbody sweep with a range of $\pm \;\qty{2}{\kelvin}$ around the mean.
The $1/f$ noise of all the KIDs is found to be significant, requiring a fit to extract the photon noise.
The fitting procedure is described in the Appendix.
One filter KID, KID 8, is found to have an abnormally high noise level due to spiking events, making a reliable estimate of the photon noise impossible.
We will therefore exclude this KID from the efficiency calculation.
For all the other KIDs we get a good fit on the noise, giving a reliable photon noise level $S_\theta$ to obtain the efficiency using \cref{eq:eta_star}.

The raw wideband data, the filter responses with wideband reference, and the relative response of the wideband KIDs after the filterbank are shown in \cref{fig:raw_response}.
The dips in the power transmitted across the filterbank are quite deep and are approximately two orders of magnitude lower than the off-resonance baseline.
These dips suggest that the filters are capturing or reflecting nearly all ($> \qty{99}{\percent}$) the power.
From the filterbank simulations in \cref{fig:simulated_fb} an on-resonance reflection $|\Gamma|$ of \qtyrange{10}{20}{\percent} is expected.
To check this, we compare the amplitudes of the wideband couplers before the filterbank around \qty{200}{\GHz}, where their relative separation is $3\lambda/4$.
Here, the wideband amplitude envelope is proportional to $1 + |\Gamma|$ for one of the wideband-coupled KIDs or $1 - |\Gamma|$ for the other, which allows us to infer $|\Gamma|$ based on their relative amplitude difference.
We find that the reflection $|\Gamma|$ varies between \qtyrange{10}{45}{\percent} around \qty{200}{\GHz}, which suggest that there is an additional reflectance above what we expect.
These additional reflectances are also visible in the filter response in \cref{fig:efficiency}a and we use the normalization method in \cref{subsec:normalization} to account for this.

Note that the two KIDs belonging to filters at 110 GHz and 135 GHz are missing, however their filter resonance dips are visible in the estimate of the power transmitted across the filterbank.
So, that must mean the aluminium part of the KIDs must be connected to the filter structure, otherwise we should see large peaks in the wideband-coupled KIDs due to reflections..
This suggests that the filters still work as intended, despite having missing KIDs.

The relative filter responses are normalized according to \cref{subsec:normalization}, resulting in the normalized filter responses $\tilde{\eta}_i(f)$ which are shown in \cref{fig:efficiency}a.
The normalized responses are stacked using a normalized frequency centered on the filter resonance.
The average shape is approximately a lorentzian and the Q-factor is equal for all filters, validating our assumptions for our normalization method.
The relatively strong ripple on the filter responses justifies the normalization method using an integrated lorentzian.

The noise and responsivity measurements are combined with the normalized filter frequency response measurement according to the methods in \cref{sec:methods}, resulting in an average measured efficiency of \qty{75}{\percent} as shown in \cref{fig:efficiency}b.
We find a $\Ql = 19.6$, which is lower than the designed Q-factor and therefore the coupling strength is stronger than we simulated.

\section{Conclusion}
We designed, fabricated, and measured a sparsely sampled filterbank using directional filters.
We measured a mean filter efficiency of \qty{75}{\percent}, well in agreement with our model calculation, where the \qty{25}{\percent} reduction in coupling compared to a perfect isolated filter can be attributed to reflections at the filters.
Besides this, we have also demonstrated a very low Q-factor of 19.6 for our on-chip filterbank using two parallel-coupled microstrip lines underneath a dielectric capping layer.

Our results show that coherent filters provide a very efficient filter solution for on-chip filterbank spectrometers, resulting in significantly higher coupling efficiency than conventional half-wave filters.
The improvement in efficiency presented in this paper makes directional filters for on-chip spectrometers the most obvious choice for scaling toward large imaging spectrometers to be installed on current and future sub-millimeter telescope facilities.

\section*{Acknowledgment}
\noindent Views and opinions expressed are those of the authors only and do not necessarily reflect those of the European Union or the European Research Council Executive Agency. Neither the European Union nor the granting authority can be held responsible for them.

\appendix

\begin{figure}[h]
    \centering
    \includegraphics[width=\linewidth]{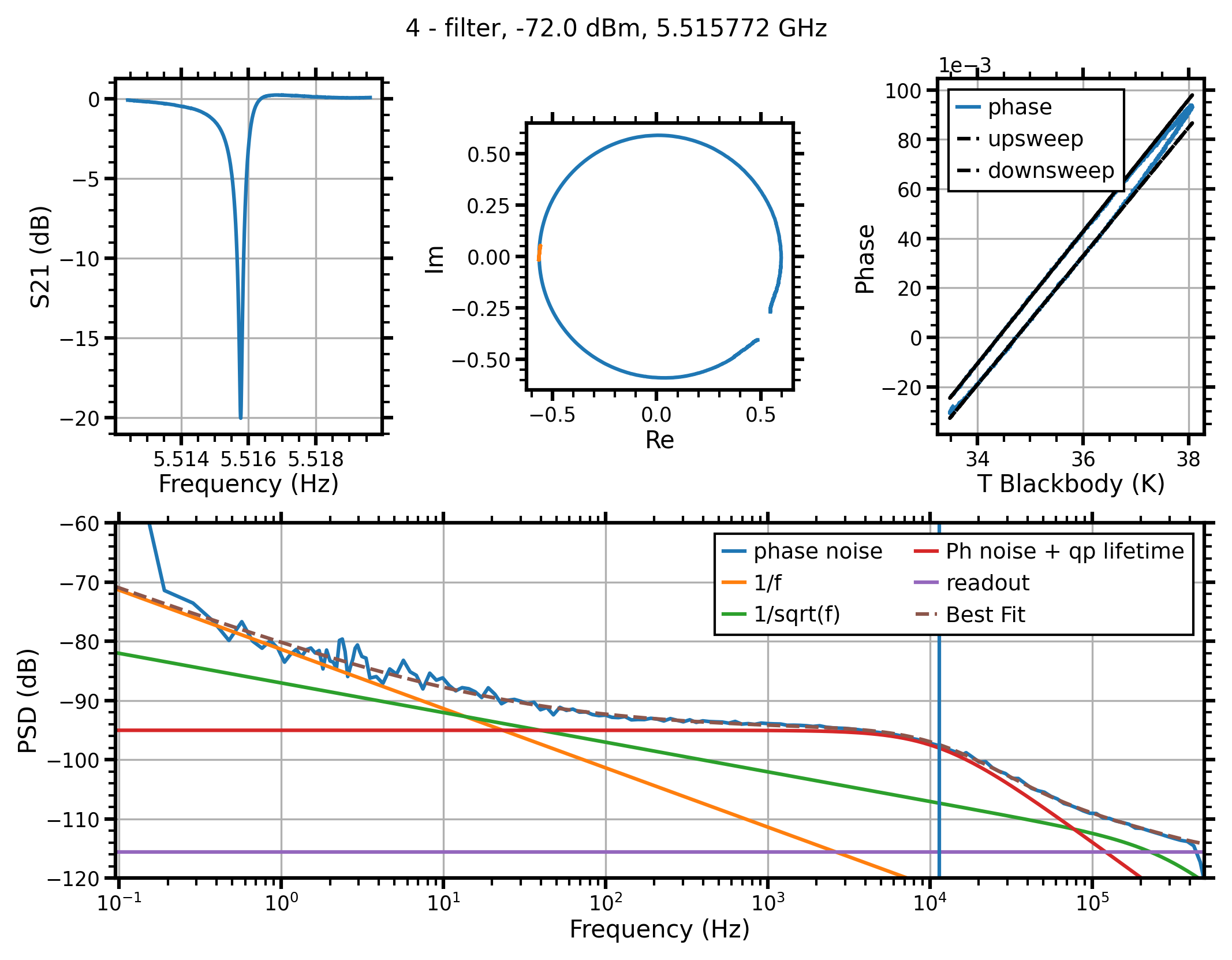}
    \caption{Results of a KID measured in the setup illustrated in \cref{fig:experiment}. The top-left plot shows the local frequency sweep of a KID. The center-top plot shows the KID circle transformed with its center at (0,0) and the measured blackbody response is shown in orange. At the top-right is the blackbody response with a fit for the upsweep and downsweep in temperature. We take the average of both for our responsivity. The hysteresis is due to the blackbody temperature lagging slightly with respect to the heater and the temperature sensor. The phase noise $S_\theta$ is shown in the bottom figure. The brown dashed line shows the best fit, with the other lines showing the decomposed components.}
    \label{fig:noise_fit}
\end{figure}

\cref{fig:noise_fit} is an example of a measured KID, showing the blackbody responsivity and a phase noise power spectral density measurement. For the responsivity to the blackbody, we take the average of the upsweep and downsweep phase responsivity in temperature. The hysteresis is due to the blackbody temperature lagging slightly with respect to the heater and the temperature sensor.

The noise is fit using a composite model with the following components:
\begin{enumerate}
	\item Photon noise with quasi-particle roll-off ($\tau_{qp}$)
	\item $1/f$ noise with resonator ring time roll-off ($\tau_{res}$)
	\item $1/\sqrt{f}$ noise with resonator ring time roll-off ($\tau_{res}$)
	\item White readout noise
\end{enumerate}
We initialize the fitting procedure by finding the flattest region between \qty{1e1}{\hertz} and \qty{1e4}{\hertz} using the minimum of the derivative of a smoothed phase noise function $S_\theta$.
The photon noise level minus 1 dB is the initial guess for the photon noise level.
The $1/f$ initial value was set at a value slightly lower than the phase noise at 1 Hz.
The $1/\sqrt{f}$ initial value was set slightly lower than the $1/f$ value, to make it sub-dominant.
During fitting, it became clear that independent $1/f$ and a $1/\sqrt{f}$ terms are needed to fit the data well.
A single $(1/f)^n$ term would not capture the slow $1/f$ to photon noise knee plus the slow photon noise roll-off to readout noise transistions simultaneously, therefore both were kept independent.
This noise decomposition and fitting procedure proved very effective at finding the photon noise level for our KIDs, since they have a significant amount of $1/f$ noise.

\printbibliography

@article{bensonSpectralCharacterizationPerformance2025,
  title = {Spectral Characterization and Performance of {{SPT-SLIM}} on-Chip Filterbank Spectrometers},
  author = {Benson, C. S. and Fichman, K. and Adamic, M. and Anderson, A. J. and Barry, P. S. and Benson, B. A. and Brooks, E. and Carlstrom, J. E. and Cecil, T. and Chang, C. L. and Dibert, K. R. and Dobbs, M. and Karkare, K. S. and Keating, G. K. and Lapuente, A. M. and Lisovenko, M. and Marrone, D. P. and Montgomery, J. and Natoli, T. and Pan, Z. and Rahlin, A. and Robson, G. and Rouble, M. and Smecher, G. and Yefremenko, V. and Young, M. R. and Yu, C. and Zebrowski, J. A. and Zhang, C.},
  year = 2025,
  month = oct,
  eprint = {2509.02245},
  primaryclass = {astro-ph},
  doi = {10.48550/arXiv.2509.02245},
  urldate = {2025-12-10},
  archiveprefix = {arXiv}
}

@inproceedings{bradfordZSpecBroadbandMillimeterwave2004,
  title = {Z-{{Spec}}, a Broadband Millimeter-Wave Grating Spectrometer--Design, Construction, and First Cryogenic Measurements},
  booktitle = {Millimeter and {{Submillimeter Detectors}} for {{Astronomy II}}},
  author = {Bradford, C. Matt and Ade, Peter A. R. and Aguirre, James E. and Bock, James J. and Dragovan, Mark and Duband, Lionel and Earle, Lieko and Glenn, Jason and Matsuhara, Hideo and Naylor, Bret J. and Nguyen, Hien T. and Yun, Minhee and Zmuidzinas, Jonas},
  year = 2004,
  month = oct,
  volume = {5498},
  pages = {257--267},
  publisher = {SPIE},
  doi = {10.1117/12.552182},
  urldate = {2025-05-15}
}

@article{buijtendorpCharacterizationLowlossHydrogenated2022,
  title = {Characterization of Low-Loss Hydrogenated Amorphous Silicon Films for Superconducting Resonators},
  author = {Buijtendorp, Bruno T. and Bueno, Juan and Thoen, David J. and Murugesan, Vignesh and Sberna, Paolo M. and Baselmans, Jochem J. A. and Vollebregt, Sten and Endo, Akira},
  year = 2022,
  month = jun,
  journal = {J. Astron. Telesc. Instrum. Syst.},
  volume = {8},
  number = {02},
  eprint = {2012.07692},
  primaryclass = {astro-ph, physics:cond-mat},
  issn = {2329-4124},
  doi = {10.1117/1.JATIS.8.2.028006},
  urldate = {2024-07-04},
  archiveprefix = {arXiv}
}

@article{cothardDesignCCATprimeEpoch2020,
  title = {The {{Design}} of the {{CCAT-prime Epoch}} of {{Reionization Spectrometer Instrument}}},
  author = {Cothard, N. F. and Choi, S. K. and Duell, C. J. and Herter, T. and Hubmayr, J. and McMahon, J. and Niemack, M. D. and Nikola, T. and Sierra, C. and Stacey, G. J. and Vavagiakis, E. M. and Wollack, E. J. and Zou, B.},
  year = 2020,
  month = may,
  journal = {J Low Temp Phys},
  volume = {199},
  number = {3},
  pages = {898--907},
  issn = {1573-7357},
  doi = {10.1007/s10909-019-02297-1},
  urldate = {2025-11-14},
  langid = {english}
}

@article{delabrouilleMicrowaveSpectroPolarimetryMatter2019,
  title = {Microwave {{Spectro-Polarimetry}} of {{Matter}} and {{Radiation}} across {{Space}} and {{Time}}},
  author = {Delabrouille, Jacques and Abitbol, Maximilian H. and Aghanim, Nabila and {Ali-Haimoud}, Yacine and Alonso, David and Alvarez, Marcelo and Banday, Anthony J. and Bartlett, James G. and Baselmans, Jochem and Basu, Kaustuv and Battaglia, Nicholas and Climent, Jose Ramon Bermejo and Bernal, Jose L. and B{\'e}thermin, Matthieu and Bolliet, Boris and Bonato, Matteo and Bouchet, Fran{\c c}ois R. and Breysse, Patrick C. and Burigana, Carlo and Cai, Zhen-Yi and Chluba, Jens and Churazov, Eugene and Dannerbauer, Helmut and De Bernardis, Paolo and De Zotti, Gianfranco and Di Valentino, Eleonora and Dimastrogiovanni, Emanuela and Endo, Akira and Erler, Jens and Ferraro, Simone and Finelli, Fabio and Fixsen, Dale and Hanany, Shaul and Hart, Luke and {Hernandez-Monteagudo}, Carlos and Hill, J. Colin and Hotinli, Selim C. and Karatsu, Kenichi and Karkare, Kirit and Keating, Garrett K. and Khabibullin, Ildar and Kogut, Alan and Kohri, Kazunori and Kovetz, Ely D. and Lagache, Guilaine and Lesgourgues, Julien and Madhavacheril, Mathew and Maffei, Bruno and Mandolesi, Nazzareno and Martins, Carlos and Masi, Silvia and Mather, John and Melin, Jean-Baptiste and Dizgah, Azadeh Moradinezhad and Mroczkowski, Tony and Mukherjee, Suvodip and Nagai, Daisuke and Negrello, Mattia and {Palanque-Delabrouille}, Nathalie and Paoletti, Daniela and Patil, Subodh P. and Piacentini, Francesco and Raghunathan, Srinivasan and Ravenni, Andrea and Remazeilles, Mathieu and Rev{\'e}ret, Vincent and Rodriguez, Louis and Rotti, Aditya and Martin, Jose-Alberto Rubino and Sayers, Jack and Scott, Douglas and Silk, Joseph and Silva, Marta and Souradeep, Tarun and Sugiyama, Naonori and Sunyaev, Rashid and Switzer, Eric R. and Tartari, Andrea and Trombetti, Tiziana and Zubeldia, Inigo},
  year = 2019,
  month = sep,
  eprint = {1909.01591},
  primaryclass = {astro-ph, physics:gr-qc},
  doi = {10.48550/arXiv.1909.01591},
  urldate = {2024-08-08},
  archiveprefix = {arXiv}
}

@article{desertContinuumCOWater2025,
  title = {Continuum, {{CO}}, and Water Vapour Maps of the {{Orion Nebula}}: {{First}} Millimetre Spectral Imaging with {{CONCERTO}}},
  shorttitle = {Continuum, {{CO}}, and Water Vapour Maps of the {{Orion Nebula}}},
  author = {D{\'e}sert, F.-X. and {Mac{\'i}as-P{\'e}rez}, J. F. and Beelen, A. and Beno{\^i}t, A. and B{\'e}thermin, M. and Bounmy, J. and Bourrion, O. and Calvo, M. and Catalano, A. and De Breuck, C. and Dubois, C. and Dur{\'a}n, C. A. and Fasano, A. and Goupy, J. and Hu, W. and Ibar, E. and Lagache, G. and Lundgren, A. and Monfardini, A. and Ponthieu, N. and Quinatoa, D. and Van Cuyck, M. and Adam, R. and Ade, P. and Ajeddig, H. and Amarantidis, S. and Andr{\'e}, P. and Aussel, H. and Berta, S. and Bongiovanni, A. and Ch{\'e}rouvrier, D. and De Petris, M. and Doyle, S. and Driessen, E. F. C. and Ejlali, G. and Ferragamo, A. and Gomez, A. and Hanser, C. and Katsioli, S. and K{\'e}ruzor{\'e}, F. and Kramer, C. and Ladjelate, B. and Leclercq, S. and Lestrade, J.-F. and Madden, S. C. and Maury, A. and Mayet, F. and {Moyer-Anin}, A. and {Mu{\~n}oz-Echeverr{\'i}a}, M. and Myserlis, I. and Paliwal, A. and Perotto, L. and Pisano, G. and Rev{\'e}ret, V. and Rigby, A. J. and Ritacco, A. and Roussel, H. and Ruppin, F. and {S{\'a}nchez-Portal}, M. and Savorgnano, S. and Sievers, A. and Tucker, C. and Zylka, R.},
  year = 2025,
  month = sep,
  journal = {A\&A},
  volume = {701},
  pages = {A210},
  issn = {0004-6361, 1432-0746},
  doi = {10.1051/0004-6361/202555320},
  urldate = {2025-11-14},
  copyright = {https://creativecommons.org/licenses/by/4.0}
}

@article{devisserFluctuationsElectronSystem2014,
  title = {Fluctuations in the Electron System of a Superconductor Exposed to a Photon Flux},
  author = {De Visser, P. J. and Baselmans, J. J. A. and Bueno, J. and Llombart, N. and Klapwijk, T. M.},
  year = 2014,
  month = feb,
  journal = {Nat Commun},
  volume = {5},
  number = {1},
  pages = {3130},
  issn = {2041-1723},
  doi = {10.1038/ncomms4130},
  urldate = {2024-07-24},
  langid = {english}
}

@article{endoFirstLightDemonstration2019,
  title = {First Light Demonstration of the Integrated Superconducting Spectrometer},
  author = {Endo, Akira and Karatsu, Kenichi and Tamura, Yoichi and Oshima, Tai and Taniguchi, Akio and Takekoshi, Tatsuya and Asayama, Shin'ichiro and Bakx, Tom J. L. C. and Bosma, Sjoerd and Bueno, Juan and Chin, Kah Wuy and Fujii, Yasunori and Fujita, Kazuyuki and Huiting, Robert and Ikarashi, Soh and Ishida, Tsuyoshi and Ishii, Shun and Kawabe, Ryohei and Klapwijk, Teun M. and Kohno, Kotaro and Kouchi, Akira and Llombart, Nuria and Maekawa, Jun and Murugesan, Vignesh and Nakatsubo, Shunichi and Naruse, Masato and Ohtawara, Kazushige and Pascual Laguna, Alejandro and Suzuki, Junya and Suzuki, Koyo and Thoen, David J. and Tsukagoshi, Takashi and Ueda, Tetsutaro and {de Visser}, Pieter J. and {van der Werf}, Paul P. and Yates, Stephen J. C. and Yoshimura, Yuki and Yurduseven, Ozan and Baselmans, Jochem J. A.},
  year = 2019,
  month = nov,
  journal = {Nat Astron},
  volume = {3},
  number = {11},
  pages = {989--996},
  publisher = {Nature Publishing Group},
  issn = {2397-3366},
  doi = {10.1038/s41550-019-0850-8},
  urldate = {2023-01-18},
  copyright = {2019 The Author(s), under exclusive licence to Springer Nature Limited},
  langid = {english}
}

@article{endoWidebandOnchipTerahertz2019,
  title = {Wideband On-Chip Terahertz Spectrometer Based on a Superconducting Filterbank},
  author = {Endo, Akira and Karatsu, Kenichi and Laguna, Alejandro Pascual and Mirzaei, Behnam and Huiting, Robert and Thoen, David and Murugesan, Vignesh and Yates, Stephen J. C. and Bueno, Juan and Marrewijk, Nuri V. and Bosma, Sjoerd and Yurduseven, Ozan and Llombart, Nuria and Suzuki, Junya and Naruse, Masato and de Visser, Pieter J. and van der Werf, Paul P. and Klapwijk, Teunis M. and Baselmans, Jochem J. A.},
  year = 2019,
  month = jun,
  journal = {JATIS},
  volume = {5},
  number = {3},
  pages = {035004},
  publisher = {SPIE},
  issn = {2329-4124, 2329-4221},
  doi = {10.1117/1.JATIS.5.3.035004},
  urldate = {2023-01-18}
}

@inproceedings{ferkinhoffDesignFirstlightPerformance2012,
  title = {Design and First-Light Performance of {{TES}} Bolometer Arrays for Submillimeter Spectroscopy with {{ZEUS-2}}},
  booktitle = {Millimeter, {{Submillimeter}}, and {{Far-Infrared Detectors}} and {{Instrumentation}} for {{Astronomy VI}}},
  author = {Ferkinhoff, Carl and Nikola, Thomas and Parshley, Stephen C. and Stacey, Gordon J. and Irwin, Kent D. and Cho, Hsiao-Mei and Niemack, Mike and Halpern, Mark and Hasselfield, Matthew and Amiri, Mandana},
  year = 2012,
  month = sep,
  volume = {8452},
  pages = {44--55},
  publisher = {SPIE},
  doi = {10.1117/12.927237},
  urldate = {2025-12-10}
}

@article{ferrariAntennaCoupledMKID2018,
  title = {Antenna {{Coupled MKID Performance Verification}} at 850 {{GHz}} for {{Large Format Astrophysics Arrays}}},
  author = {Ferrari, Lorenza and Yurduseven, Ozan and Llombart, Nuria and Yates, Stephen J. C. and Bueno, Juan and Murugesan, Vignesh and Thoen, David J. and Endo, Akira and Baryshev, Andrey M. and Baselmans, Jochem J. A.},
  year = 2018,
  month = jan,
  journal = {IEEE Transactions on Terahertz Science and Technology},
  volume = {8},
  number = {1},
  pages = {127--139},
  issn = {2156-3446},
  doi = {10.1109/TTHZ.2017.2764378},
  urldate = {2024-05-14}
}

@article{fronenbergForecastsStatisticalInsights2024,
  title = {Forecasts and {{Statistical Insights}} for {{Line Intensity Mapping Cross-correlations}}: {{A Case Study}} with 21 Cm \texttimes{} [{{C}} Ii]},
  shorttitle = {Forecasts and {{Statistical Insights}} for {{Line Intensity Mapping Cross-correlations}}},
  author = {Fronenberg, Hannah and Liu, Adrian},
  year = 2024,
  month = nov,
  journal = {ApJ},
  volume = {975},
  number = {2},
  pages = {222},
  publisher = {The American Astronomical Society},
  issn = {0004-637X},
  doi = {10.3847/1538-4357/ad77cc},
  urldate = {2025-10-28},
  langid = {english}
}

@article{hahnleSuperconductingMicrostripLosses2021,
  title = {Superconducting {{Microstrip Losses}} at {{Microwave}} and {{Submillimeter Wavelengths}}},
  author = {H{\"a}hnle, S. and Kouwenhoven, K. and Buijtendorp, B. and Endo, A. and Karatsu, K. and Thoen, D.J. and Murugesan, V. and Baselmans, J.J.A.},
  year = 2021,
  month = jul,
  journal = {Phys. Rev. Appl.},
  volume = {16},
  number = {1},
  pages = {014019},
  publisher = {American Physical Society},
  doi = {10.1103/PhysRevApplied.16.014019},
  urldate = {2024-11-18}
}

@article{karoumpisCIILineIntensity2024,
  title = {[{{CII}}] Line Intensity Mapping the Epoch of Reionization with the {{Prime-Cam}} on {{FYST}} - {{II}}. {{CO}} Foreground Masking Based on an External Catalog},
  author = {Karoumpis, C. and Magnelli, B. and {Romano-D{\'i}az}, E. and Garcia, K. and Dev, A. and Clarke, J. and Wang, T.-M. and B{\u a}descu, T. and Riechers, D. and Bertoldi, F.},
  year = 2024,
  month = nov,
  journal = {A\&A},
  volume = {691},
  pages = {A262},
  publisher = {EDP Sciences},
  issn = {0004-6361, 1432-0746},
  doi = {10.1051/0004-6361/202450304},
  urldate = {2025-10-27},
  copyright = {\copyright{} ESO 2024},
  langid = {english}
}

@article{marcuzzoConstrainingIILuminosity2025,
  title = {Constraining the [{{C II}}] Luminosity Function from the Power Spectrum of Line-Intensity Maps at Redshift 3.6},
  author = {Marcuzzo, Elena and Porciani, Cristiano and {Romano-D{\'i}az}, Emilio and Khatri, Prachi},
  year = 2025,
  month = aug,
  journal = {A\&A},
  volume = {700},
  pages = {A211},
  publisher = {EDP Sciences},
  issn = {0004-6361, 1432-0746},
  doi = {10.1051/0004-6361/202554970},
  urldate = {2025-10-27},
  copyright = {\copyright{} The Authors 2025},
  langid = {english}
}

@article{martingDirectionalFilterDesign2024a,
  title = {Directional {{Filter Design}} and {{Simulation}} for {{Superconducting On-Chip Filter-Banks}}},
  author = {Marting, Louis H. and Karatsu, Kenichi and Endo, Akira and Baselmans, Jochem J. A. and Pascual Laguna, Alejandro},
  year = 2024,
  month = may,
  journal = {J Low Temp Phys},
  issn = {1573-7357},
  doi = {10.1007/s10909-024-03118-w},
  urldate = {2024-05-06},
  langid = {english}
}

@inproceedings{mirzaeiUspecSpectrometersEXCLAIM2020,
  title = {\textmu -Spec Spectrometers for the {{EXCLAIM}} Instrument},
  booktitle = {Millimeter, {{Submillimeter}}, and {{Far-Infrared Detectors}} and {{Instrumentation}} for {{Astronomy X}}},
  author = {Mirzaei, Mona and Barrentine, Emily M. and Bulcha, Berhanu T. and Cataldo, Giuseppe and Connors, Jake A. and Ehsan, Negar and {Essinger-Hileman}, Thomas M. and Hess, Larry A. and {Mugge-Durum}, Jonas W. and Noroozian, Omid and Oxholm, Trevor M. and Stevenson, Thomas R. and Switzer, Eric R. and Volpert, Carolyn G. and Wollack, Edward J.},
  year = 2020,
  month = dec,
  volume = {11453},
  pages = {128--139},
  publisher = {SPIE},
  doi = {10.1117/12.2562446},
  urldate = {2024-08-20}
}

@article{mroczkowskiAstrophysicsSpatiallySpectrally2019,
  title = {Astrophysics with the {{Spatially}} and {{Spectrally Resolved Sunyaev-Zeldovich Effects}}},
  author = {Mroczkowski, Tony and Nagai, Daisuke and Basu, Kaustuv and Chluba, Jens and Sayers, Jack and Adam, R{\'e}mi and Churazov, Eugene and Crites, Abigail and Di Mascolo, Luca and Eckert, Dominique and {Macias-Perez}, Juan and Mayet, Fr{\'e}d{\'e}ric and Perotto, Laurence and Pointecouteau, Etienne and Romero, Charles and Ruppin, Florian and Scannapieco, Evan and ZuHone, John},
  year = 2019,
  month = feb,
  journal = {Space Sci Rev},
  volume = {215},
  number = {1},
  pages = {17},
  issn = {1572-9672},
  doi = {10.1007/s11214-019-0581-2},
  urldate = {2024-06-10},
  langid = {english}
}

@article{mroczkowskiConceptualDesign50meter2025,
  title = {The Conceptual Design of the 50-Meter {{Atacama Large Aperture Submillimeter Telescope}} ({{AtLAST}})},
  author = {Mroczkowski, Tony and Gallardo, Patricio A. and Timpe, Martin and Kiselev, Aleksej and Groh, Manuel and Kaercher, Hans and Reichert, Matthias and Cicone, Claudia and Puddu, Roberto and {Dubois-dit-Bonclaude}, Pierre and Bok, Daniel and Dahl, Erik and Macintosh, Mike and Dicker, Simon and Viole, Isabelle and Sartori, Sabrina and Venegas, Guillermo Andr{\'e}s Valenzuela and Zeyringer, Marianne and Niemack, Michael and Poppi, Sergio and Olguin, Rodrigo and Hatziminaoglou, Evanthia and Breuck, Carlos De and Klaassen, Pamela and {Montenegro-Montes}, Francisco Miguel and Zimmerer, Thomas},
  year = 2025,
  month = feb,
  journal = {A\&A},
  volume = {694},
  pages = {A142},
  publisher = {EDP Sciences},
  issn = {0004-6361, 1432-0746},
  doi = {10.1051/0004-6361/202449786},
  urldate = {2025-10-27},
  copyright = {\copyright{} The Authors 2025},
  langid = {english}
}

@inproceedings{parshleyCCATprimeFredYoung2022,
  title = {{{CCAT-prime}}: The {{Fred Young Submillimeter Telescope}} ({{FYST}}) Final Design and Fabrication},
  shorttitle = {{{CCAT-prime}}},
  booktitle = {Ground-Based and {{Airborne Telescopes IX}}},
  author = {Parshley, Stephen C. and Gramke, Scott and Higgins, Ronan and Kronshage, J{\"o}rg and Steeger, Karl and Willmeroth, Klaus},
  year = 2022,
  month = aug,
  volume = {12182},
  pages = {543--550},
  publisher = {SPIE},
  doi = {10.1117/12.2630515},
  urldate = {2025-12-10}
}

@article{pascuallagunaTerahertzBandPassFilters2021,
  title = {Terahertz {{Band-Pass Filters}} for {{Wideband Superconducting On-Chip Filter-Bank Spectrometers}}},
  author = {Pascual Laguna, Alejandro and Karatsu, Kenichi and Thoen, David J. and Murugesan, Vignesh and Buijtendorp, Bruno T. and Endo, Akira and Baselmans, Jochem J. A.},
  year = 2021,
  month = nov,
  journal = {IEEE Transactions on Terahertz Science and Technology},
  volume = {11},
  number = {6},
  pages = {635--646},
  issn = {2156-3446},
  doi = {10.1109/TTHZ.2021.3095429}
}

@article{redfordSuperSpecOnChipSpectrometer2022,
  title = {{{SuperSpec}}: {{On-Chip Spectrometer Design}}, {{Characterization}}, and {{Performance}}},
  shorttitle = {{{SuperSpec}}},
  author = {Redford, J. and Barry, P. S. and Bradford, C. M. and Chapman, S. and Glenn, J. and {Hailey-Dunsheath}, S. and Janssen, R. M. J. and Karkare, K. S. and LeDuc, H. G. and Mauskopf, P. and McGeehan, R. and Shirokoff, E. and Wheeler, J. and Zmuidzinas, J.},
  year = 2022,
  month = nov,
  journal = {J Low Temp Phys},
  volume = {209},
  number = {3},
  pages = {548--555},
  issn = {1573-7357},
  doi = {10.1007/s10909-022-02866-x},
  urldate = {2023-08-30},
  langid = {english}
}

@article{royCrosscorrelationTechniquesMitigate2024,
  title = {Cross-Correlation {{Techniques}} to {{Mitigate}} the {{Interloper Contamination}} for {{Line Intensity Mapping Experiments}}},
  author = {Roy, Anirban and Battaglia, Nicholas},
  year = 2024,
  month = jun,
  journal = {ApJ},
  volume = {969},
  number = {1},
  pages = {2},
  publisher = {The American Astronomical Society},
  issn = {0004-637X},
  doi = {10.3847/1538-4357/ad4a29},
  urldate = {2025-10-27},
  langid = {english}
}

@article{sptpolcollaborationDetectionBModePolarization2013,
  title = {Detection of {{B-Mode Polarization}} in the {{Cosmic Microwave Background}} with {{Data}} from the {{South Pole Telescope}}},
  author = {{SPTpol Collaboration} and Hanson, D. and Hoover, S. and Crites, A. and Ade, P. A. R. and Aird, K. A. and Austermann, J. E. and Beall, J. A. and Bender, A. N. and Benson, B. A. and Bleem, L. E. and Bock, J. J. and Carlstrom, J. E. and Chang, C. L. and Chiang, H. C. and Cho, H-M. and Conley, A. and Crawford, T. M. and {de Haan}, T. and Dobbs, M. A. and Everett, W. and Gallicchio, J. and Gao, J. and George, E. M. and Halverson, N. W. and Harrington, N. and Henning, J. W. and Hilton, G. C. and Holder, G. P. and Holzapfel, W. L. and Hrubes, J. D. and Huang, N. and Hubmayr, J. and Irwin, K. D. and Keisler, R. and Knox, L. and Lee, A. T. and Leitch, E. and Li, D. and Liang, C. and {Luong-Van}, D. and Marsden, G. and McMahon, J. J. and Mehl, J. and Meyer, S. S. and Mocanu, L. and Montroy, T. E. and Natoli, T. and Nibarger, J. P. and Novosad, V. and Padin, S. and Pryke, C. and Reichardt, C. L. and Ruhl, J. E. and Saliwanchik, B. R. and Sayre, J. T. and Schaffer, K. K. and Schulz, B. and Smecher, G. and Stark, A. A. and Story, K. T. and Tucker, C. and Vanderlinde, K. and Vieira, J. D. and Viero, M. P. and Wang, G. and Yefremenko, V. and Zahn, O. and Zemcov, M.},
  year = 2013,
  month = sep,
  journal = {Phys. Rev. Lett.},
  volume = {111},
  number = {14},
  pages = {141301},
  publisher = {American Physical Society},
  doi = {10.1103/PhysRevLett.111.141301},
  urldate = {2025-10-16}
}

@article{taniguchiDESHIMA20Development2022,
  title = {{{DESHIMA}} 2.0: {{Development}} of an {{Integrated Superconducting Spectrometer}} for {{Science-Grade Astronomical Observations}}},
  shorttitle = {{{DESHIMA}} 2.0},
  author = {Taniguchi, Akio and Bakx, Tom J. L. C. and Baselmans, Jochem J. A. and Huiting, Robert and Karatsu, Kenichi and Llombart, Nuria and Rybak, Matus and Takekoshi, Tatsuya and Tamura, Yoichi and Akamatsu, Hiroki and Brackenhoff, Stefanie and Bueno, Juan and Buijtendorp, Bruno T. and Dabironezare, Shahab O. and Doing, Anne-Kee and Fujii, Yasunori and Fujita, Kazuyuki and Gouwerok, Matthijs and H{\"a}hnle, Sebastian and Ishida, Tsuyoshi and Ishii, Shun and Kawabe, Ryohei and Kitayama, Tetsu and Kohno, Kotaro and Kouchi, Akira and Maekawa, Jun and Matsuda, Keiichi and Murugesan, Vignesh and Nakatsubo, Shunichi and Oshima, Tai and Pascual Laguna, Alejandro and Thoen, David J. and {van der Werf}, Paul P. and Yates, Stephen J. C. and Endo, Akira},
  year = 2022,
  month = nov,
  journal = {J Low Temp Phys},
  volume = {209},
  number = {3},
  pages = {278--286},
  issn = {1573-7357},
  doi = {10.1007/s10909-022-02888-5},
  urldate = {2023-10-12},
  langid = {english}
}

@article{theconcertocollaborationWideFieldofviewLowresolution2020,
  title = {A Wide Field-of-View Low-Resolution Spectrometer at {{APEX}}: {{Instrument}} Design and Scientific Forecast},
  shorttitle = {A Wide Field-of-View Low-Resolution Spectrometer at {{APEX}}},
  author = {{The CONCERTO Collaboration} and Ade, P. and Aravena, M. and Barria, E. and Beelen, A. and Benoit, A. and B{\'e}thermin, M. and Bounmy, J. and Bourrion, O. and Bres, G. and De Breuck, C. and Calvo, M. and Cao, Y. and Catalano, A. and D{\'e}sert, F.-X. and Dur{\'a}n, C.A. and Fasano, A. and Fenouillet, T. and Garcia, J. and Garde, G. and Goupy, J. and Groppi, C. and Hoarau, C. and Lagache, G. and Lambert, J.-C. and Leggeri, J.-P. and {Levy-Bertrand}, F. and {Mac{\'i}as-P{\'e}rez}, J. and Mani, H. and Marpaud, J. and Mauskopf, P. and Monfardini, A. and Pisano, G. and Ponthieu, N. and Prieur, L. and Roni, S. and Roudier, S. and Tourres, D. and Tucker, C.},
  year = 2020,
  month = oct,
  journal = {A\&A},
  volume = {642},
  pages = {A60},
  issn = {0004-6361, 1432-0746},
  doi = {10.1051/0004-6361/202038456},
  urldate = {2025-11-14},
  copyright = {https://creativecommons.org/licenses/by/4.0}
}

@article{thoenCombinedUltravioletElectronbeam2022,
  title = {Combined Ultraviolet- and Electron-Beam Lithography with {{Micro-Resist-Technology GmbH}} Ma-{{N1400}} Resist},
  author = {Thoen, D. J. and Murugesan, V. and Pascual Laguna, A. and Karatsu, K. and Endo, A. and Baselmans, J. J. A.},
  year = 2022,
  month = sep,
  journal = {J. Vac. Sci. Technol. B},
  volume = {40},
  number = {5},
  pages = {052603},
  issn = {2166-2746},
  doi = {10.1116/6.0001918},
  urldate = {2025-11-14}
}

@article{thoenSuperconductingNbTinThin2017,
  title = {Superconducting {{NbTin Thin Films With Highly Uniform Properties Over}} a \textbackslash varnothing 100 Mm {{Wafer}}},
  author = {Thoen, David Johannes and Bos, Boy Gustaaf Cornelis and Haalebos, E. A. F. and Klapwijk, T. M. and Baselmans, J. J. A. and Endo, Akira},
  year = 2017,
  month = jun,
  journal = {IEEE Transactions on Applied Superconductivity},
  volume = {27},
  number = {4},
  pages = {1--5},
  issn = {1558-2515},
  doi = {10.1109/TASC.2016.2631948},
  urldate = {2024-04-30}
}

@article{vankampenAtacamaLargeAperture2024a,
  title = {Atacama {{Large Aperture Submillimeter Telescope}} ({{AtLAST}}) Science: {{Surveying}} the Distant {{Universe}}},
  shorttitle = {Atacama {{Large Aperture Submillimeter Telescope}} ({{AtLAST}}) Science},
  author = {{van Kampen}, Eelco and Bakx, Tom and De Breuck, Carlos and Chen, Chian-Chou and Dannerbauer, Helmut and Magnelli, Benjamin and {Montenegro-Montes}, Francisco Miguel and Okumura, Teppei and Pu, Sy-Yin and Rybak, Matus and Saintonge, Amelie and Cicone, Claudia and Hatziminaoglou, Evanthia and Hilhorst, Juli{\"e}tte and Klaassen, Pamela and Lee, Minju and Lovell, Christopher C. and Lundgren, Andreas and Di Mascolo, Luca and Mroczkowski, Tony and Sommovigo, Laura and Booth, Mark and Cordiner, Martin A. and Ivison, Rob and Johnstone, Doug and Liu, Daizhong and Maccarone, Thomas J. and Smith, Matthew and Thelen, Alexander E. and Wedemeyer, Sven},
  year = 2024,
  month = jun,
  journal = {Open Res Eur},
  volume = {4},
  pages = {122},
  issn = {2732-5121},
  doi = {10.12688/openreseurope.17445.1},
  urldate = {2025-10-27},
  pmcid = {PMC11472272},
  pmid = {39403450}
}

@article{vanrantwijkMultiplexedReadout1000Pixel2016,
  title = {Multiplexed {{Readout}} for 1000-{{Pixel Arrays}} of {{Microwave Kinetic Inductance Detectors}}},
  author = {Van Rantwijk, Joris and Grim, Martin and Van Loon, Dennis and Yates, Stephen and Baryshev, Andrey and Baselmans, Jochem},
  year = 2016,
  month = jun,
  journal = {IEEE Trans. Microwave Theory Techn.},
  volume = {64},
  number = {6},
  pages = {1876--1883},
  issn = {0018-9480, 1557-9670},
  doi = {10.1109/TMTT.2016.2544303},
  urldate = {2024-07-24}
}

@article{youngDesignPerformanceSPTSLIM2025a,
  title = {Design and {{Performance}} of the {{SPT-SLIM Receiver Cryostat}}},
  author = {Young, M. R. and Adamic, M. and Anderson, A. J. and Barry, P. S. and Benson, B. A. and Benson, C. S. and Brooks, E. and Carlstrom, J. E. and Cecil, T. and Chang, C. L. and Dibert, K. R. and Dobbs, M. and Fichman, K. and Hollister, M. and Karkare, K. S. and Keating, G. K. and Lapuente, A. M. and Lisovenko, M. and Marrone, D. P. and Mitchell, D. and Montgomery, J. and Natoli, T. and Pan, Z. and Rahlin, A. and Robson, G. and Rouble, M. and Smecher, G. and Yefremenko, V. and Yu, C. and Zebrowski, J. A. and Zhang, C.},
  year = 2025,
  month = oct,
  eprint = {2510.14219},
  primaryclass = {astro-ph},
  doi = {10.48550/arXiv.2510.14219},
  urldate = {2025-12-11},
  archiveprefix = {arXiv}
}

@article{buijtendorpVibrationalModesOrigin2025,
  title = {Vibrational Modes as the Origin of Dielectric Loss at 0.27--100 {{THz}} in a-{{SiC}}:{{H}}},
  shorttitle = {Vibrational Modes as the Origin of Dielectric Loss at 0.27--100 {{THz}} in a - {{Si C}}},
  author = {Buijtendorp, B.T. and Endo, A. and Jellema, W. and Karatsu, K. and Kouwenhoven, K. and Lamers, D. and Van Der Linden, A.J. and Rostem, K. and Veen, H.M. and Wollack, E.J. and Baselmans, J.J.A. and Vollebregt, S.},
  year = 2025,
  month = jan,
  journal = {Phys Rev Appl},
  volume = {23},
  number = {1},
  pages = {014035},
  issn = {2331-7019},
  doi = {10.1103/PhysRevApplied.23.014035},
  urldate = {2026-02-18},
  langid = {english}
}

@article{dibertOnSkyAtmosphericCalibration2025,
  title = {An {{On-Sky Atmospheric Calibration}} of {{SPT-SLIM}}},
  author = {Dibert, K. R. and Adamic, M. and Anderson, A. J. and Barry, P. S. and Benson, B. A. and Benson, C. S. and Brooks, E. and Carlstrom, J. E. and Cecil, T. and Chang, C. L. and Dobbs, M. and Fichman, K. and Karkare, K. S. and Keating, G. K. and Lapuente, A. M. and Lisovenko, M. and Marrone, D. P. and Montgomery, J. and Natoli, T. and Pan, Z. and Rahlin, A. and Robson, G. and Rouble, M. and Smecher, G. and Yefremenko, V. and Young, M. R. and Yu, C. and Zebrowski, J. A. and Zhang, C.},
  year = 2025,
  month = oct,
  eprint = {2510.14220},
  primaryclass = {astro-ph},
  publisher = {arXiv},
  doi = {10.48550/arXiv.2510.14220},
  urldate = {2026-02-09},
  archiveprefix = {arXiv}
}

@article{dimascoloAtacamaLargeAperture2025,
  title = {Atacama {{Large Aperture Submillimeter Telescope}} ({{AtLAST}}) Science: {{Resolving}} the Hot and Ionized {{Universe}} through the {{Sunyaev-Zeldovich}} Effect},
  shorttitle = {Atacama {{Large Aperture Submillimeter Telescope}} ({{AtLAST}}) Science},
  author = {Di Mascolo, Luca and Perrott, Yvette and Mroczkowski, Tony and Raghunathan, Srinivasan and Andreon, Stefano and Ettori, Stefano and Simionescu, Aurora and Van Marrewijk, Joshiwa and Cicone, Claudia and Lee, Minju and Nelson, Dylan and Sommovigo, Laura and Booth, Mark and Klaassen, Pamela and Andreani, Paola and Cordiner, Martin A. and Johnstone, Doug and Van Kampen, Eelco and Liu, Daizhong and Maccarone, Thomas J. and Morris, Thomas W. and {Orlowski-Scherer}, John and Saintonge, Am{\'e}lie and Smith, Matthew and Thelen, Alexander E. and Wedemeyer, Sven},
  year = 2025,
  month = jun,
  journal = {Open Res Eur},
  volume = {4},
  pages = {113},
  issn = {2732-5121},
  doi = {10.12688/openreseurope.17449.2},
  urldate = {2026-02-18},
  langid = {english}
}

@article{huCONCERTOAPEXOnsky2024,
  title = {{{CONCERTO}} at {{APEX On-sky}} Performance in Continuum},
  author = {Hu, W. and Beelen, A. and Lagache, G. and Fasano, A. and Lundgren, A. and Ade, P. and Aravena, M. and Barria, E. and Benoit, A. and B{\'e}thermin, M. and Bounmy, J. and Bourrion, O. and Bres, G. and Breuck, C. De and Calvo, M. and Catalano, A. and D{\'e}sert, F.-X. and Dubois, C. and Dur{\'a}n, C. A. and Fenouillet, T. and Garcia, J. and Garde, G. and Goupy, J. and Hoarau, C. and Lambert, J.-C. and Lellouch, E. and {Levy-Bertrand}, F. and {Macias-Perez}, J. and Marpaud, J. and Monfardini, A. and Pisano, G. and Ponthieu, N. and Prieur, L. and Quinatoa, D. and Roni, S. and Roudier, S. and Tourres, D. and Tucker, C. and Cuyck, M. Van},
  year = 2024,
  month = sep,
  journal = {Astron Astrophys},
  volume = {689},
  pages = {A20},
  publisher = {EDP Sciences},
  issn = {0004-6361, 1432-0746},
  doi = {10.1051/0004-6361/202449260},
  urldate = {2024-12-04},
  copyright = {\copyright{} The Authors 2024},
  langid = {english}
}

@article{jovanovic2023AstrophotonicsRoadmap2023,
  title = {2023 {{Astrophotonics Roadmap}}: Pathways to Realizing Multi-Functional Integrated Astrophotonic Instruments},
  shorttitle = {2023 {{Astrophotonics Roadmap}}},
  author = {Jovanovic, Nemanja and Gatkine, Pradip and Anugu, Narsireddy and {Amezcua-Correa}, Rodrigo and Basu Thakur, Ritoban and Beichman, Charles and Bender, Chad F. and Berger, Jean-Philippe and Bigioli, Azzurra and {Bland-Hawthorn}, Joss and Bourdarot, Guillaume and Bradford, Charles M and Broeke, Ronald and Bryant, Julia and Bundy, Kevin and Cheriton, Ross and Cvetojevic, Nick and Diab, Momen and Diddams, Scott A and Dinkelaker, Aline N and Duis, Jeroen and Eikenberry, Stephen and Ellis, Simon and Endo, Akira and Figer, Donald F and Fitzgerald, Michael P. and {Gris-Sanchez}, Itandehui and Gross, Simon and Grossard, Ludovic and Guyon, Olivier and Haffert, Sebastiaan Y and Halverson, Samuel and Harris, Robert J and He, Jinping and Herr, Tobias and Hottinger, Philipp and Huby, Elsa and Ireland, Michael and {Jenson-Clem}, Rebecca and Jewell, Jeffrey and Jocou, Laurent and Kraus, Stefan and Labadie, Lucas and Lacour, Sylvestre and Laugier, Romain and {\L}awniczuk, Katarzyna and Lin, Jonathan and Leifer, Stephanie and {Leon-Saval}, Sergio and Martin, Guillermo and Martinache, Frantz and Martinod, Marc-Antoine and Mazin, Benjamin A and Minardi, Stefano and Monnier, John D and Moreira, Reinan and Mourard, Denis and Nayak, Abani Shankar and Norris, Barnaby and Obrzud, Ewelina and Perraut, Karine and Reynaud, Fran{\c c}ois and Sallum, Steph and Schiminovich, David and Schwab, Christian and Serbayn, Eugene and Soliman, Sherif and Stoll, Andreas and Tang, Liang and Tuthill, Peter and Vahala, Kerry and Vasisht, Gautam and Veilleux, Sylvain and Walter, Alexander B and Wollack, Edward J and Xin, Yinzi and Yang, Zongyin and Yerolatsitis, Stephanos and Zhang, Yang and Zou, Chang-Ling},
  year = 2023,
  month = oct,
  journal = {J Phys Photonics},
  volume = {5},
  number = {4},
  pages = {042501},
  publisher = {IOP Publishing},
  issn = {2515-7647},
  doi = {10.1088/2515-7647/ace869},
  urldate = {2026-02-09},
  langid = {english}
}

@article{karatsuDESHIMA202004002026,
  title = {{{DESHIMA}} 2.0: {{A}} 200-400 {{GHz Ultra-wideband Integrated Superconducting Spectrometer}}},
  shorttitle = {{{DESHIMA}} 2.0},
  author = {Karatsu, K. and Endo, A. and Moerman, A. and Yates, S. J. C. and Huiting, R. and Laguna, A. Pascual and Dabironezare, S. and Murugesan, V. and Thoen, D. J. and Buijtendorp, B. T. and Cray, S. and Fujita, K. and H{\"a}hnle, S. and Hanany, S. and Kawabe, R. and Kohno, K. and Marting, L. H. and Matsumura, T. and Nakatsubo, S. and Scholtenhuis, L. G. G. Olde and Oshima, T. and Rybak, M. and Steenvoorde, F. and Takaku, R. and Takekoshi, T. and Tamura, Y. and Taniguchi, A. and van der Werf, P. P. and Baselmans, J. J. A.},
  year = 2026,
  month = jan,
  eprint = {2601.21603},
  primaryclass = {astro-ph},
  publisher = {arXiv},
  doi = {10.48550/arXiv.2601.21603},
  urldate = {2026-02-02},
  archiveprefix = {arXiv}
}

@inproceedings{kohnoLargeFormatImaging2020,
  title = {Large Format Imaging Spectrograph for the {{Large Submillimeter Telescope}} ({{LST}})},
  booktitle = {Millimeter, {{Submillimeter}}, and {{Far-Infrared Detectors}} and {{Instrumentation}} for {{Astronomy X}}},
  author = {Kohno, K. and Kawabe, R. and Tamura, Y. and Endo, A. and Baselmans, J. J. A. and Karatsu, K. and Inoue, A. K. and Moriwaki, K. and Hayatsu, N. H. and Yoshida, N. and Yoshimura, Y. and Hatsukade, B. and Umehata, H. and Oshima, T. and Takekoshi, T. and Taniguchi, A. and Klaassen, P. D. and Mroczkowski, T. and Cicone, C. and Bertoldi, F. and Dannerbauer, H. and Tosaki, T.},
  year = 2020,
  month = dec,
  eprint = {2102.08280},
  primaryclass = {astro-ph},
  pages = {24},
  doi = {10.1117/12.2561238},
  urldate = {2026-02-09},
  archiveprefix = {arXiv}
}

@article{kovacsConceptIntegralField2025,
  title = {Concept Integral Field Unit Spectrometer Instrument for the Next-Generation Millimeter-Wave Cosmological Surveys},
  author = {Kov{\'a}cs, Attila and Keating, Garrett K. and Greve, Thomas and Norton, Timothy},
  year = 2025,
  month = dec,
  journal = {J Astron Telesc Instrum Syst},
  volume = {11},
  number = {4},
  pages = {045007},
  publisher = {SPIE},
  issn = {2329-4124, 2329-4221},
  doi = {10.1117/1.JATIS.11.4.045007},
  urldate = {2026-02-18}
}

@book{pozarMicrowaveEngineering2012,
  title = {Microwave Engineering},
  author = {Pozar, David M.},
  year = 2012,
  edition = {4th ed},
  publisher = {Wiley},
  address = {Hoboken, NJ},
  isbn = {978-0-470-63155-3},
  lccn = {TK7876 .P69 2012}
}

@article{robsonSimulationDesignOnChip2022,
  title = {The {{Simulation}} and {{Design}} of an {{On-Chip Superconducting Millimetre Filter-Bank Spectrometer}}},
  author = {Robson, G. and Anderson, A. J. and Barry, P. S. and Doyle, S. and Karkare, K. S.},
  year = 2022,
  month = nov,
  journal = {J Low Temp Phys},
  volume = {209},
  number = {3-4},
  pages = {493--501},
  issn = {0022-2291, 1573-7357},
  doi = {10.1007/s10909-022-02747-3},
  urldate = {2025-10-28},
  langid = {english}
}

@article{yatesSurfaceWaveControl2017,
  title = {Surface {{Wave Control}} for {{Large Arrays}} of {{Microwave Kinetic Inductance Detectors}}},
  author = {Yates, Stephen J. C. and Baryshev, Andrey M. and Yurduseven, Ozan and Bueno, Juan and Davis, Kristina K. and Ferrari, Lorenza and Jellema, Willem and Llombart, Nuria and Murugesan, Vignesh and Thoen, David J. and Baselmans, Jochem J. A.},
  year = 2017,
  month = nov,
  journal = {IEEE Trans Terahertz Sci Technol},
  volume = {7},
  number = {6},
  pages = {789--799},
  issn = {2156-3446},
  doi = {10.1109/TTHZ.2017.2755500},
  urldate = {2026-01-28}
}

@article{zhangFourierOpticsTool2021,
  title = {A {{Fourier Optics Tool}} to {{Derive}} the {{Plane Wave Spectrum}} of {{Quasi-Optical Systems}} [{{EM Programmer}}'s {{Notebook}}]},
  author = {Zhang, Huasheng and Dabironezare, Shahab Oddin and Carluccio, Giorgio and Neto, Andrea and Llombart, Nuria},
  year = 2021,
  month = feb,
  journal = {IEEE Trans Antennas Propag},
  volume = {63},
  number = {1},
  pages = {103--116},
  issn = {1558-4143},
  doi = {10.1109/MAP.2020.3027233},
  urldate = {2026-01-29}
}

\begin{IEEEbiography}[{\includegraphics[width=1in,height=1.25in,
clip,keepaspectratio]{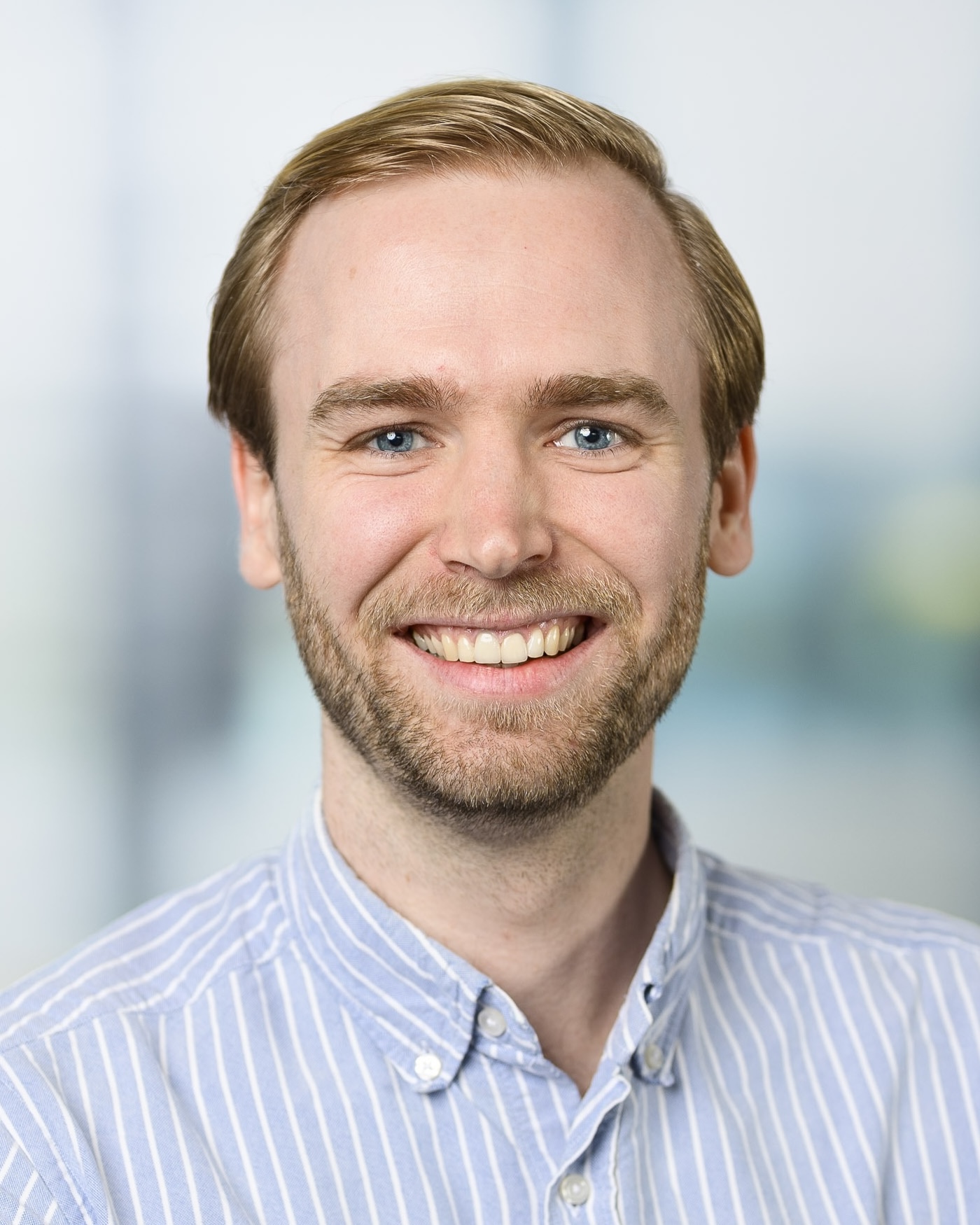}}]{Louis H. Marting}
received the B.Sc. and M.Sc. degrees (\textit{cum laude}) in electrical engineering from the Delft University of Technology, Delft, The Netherlands (TU Delft), in 2018 and 2022, respectively.
His master thesis was carried out with the Terahertz Sensing Group, TU Delft, where he is currently working toward the Ph.D. degree in electrical engineering.

During his masters degree, he specialized in on-chip RF applications, with an emphasis on superconducting on-chip spectrometers for astronomical applications at sub-mm wavelengths. This work he is continuing in his Ph.D. degree for the Terahertz Integrated Field Units with Universal Nanotechnology (TIFUUN) project.
\end{IEEEbiography}

\begin{IEEEbiography}[{\includegraphics[width=1in,height=1.25in,
clip,keepaspectratio]{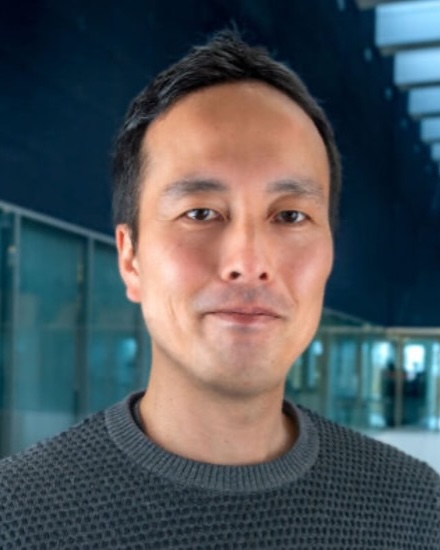}}]{Kenichi Karatsu}
received Ph.D. degree in science from Kyoto University in 2011.
His doctoral dissertations was entitled “Measurement of Cross Section and Single Spin Asymmetries of W\^{}\{+/-\} Boson Production in Polarized pp Collisions at sqrt\{s\} = 500 GeV”.

In 2011, he joined Advanced Technology Center (ATC) of National Astronomical Observatory of Japan (NAOJ) as a Postdoctoral Researcher. Since 2015, he has been working on DESHIMA project at Delft University of Technology/ SRON Netherlands Institute for Space Research. He is currently an Instrument Scientist with SRON.
His main role is to lead laboratory evaluation and telescope campaign of the instrument. His research interest is to develop an experimental instrument for revealing mysteries of the Universe.
\end{IEEEbiography}

\begin{IEEEbiography}[{\includegraphics[width=1in,height=1.25in,
clip,keepaspectratio]{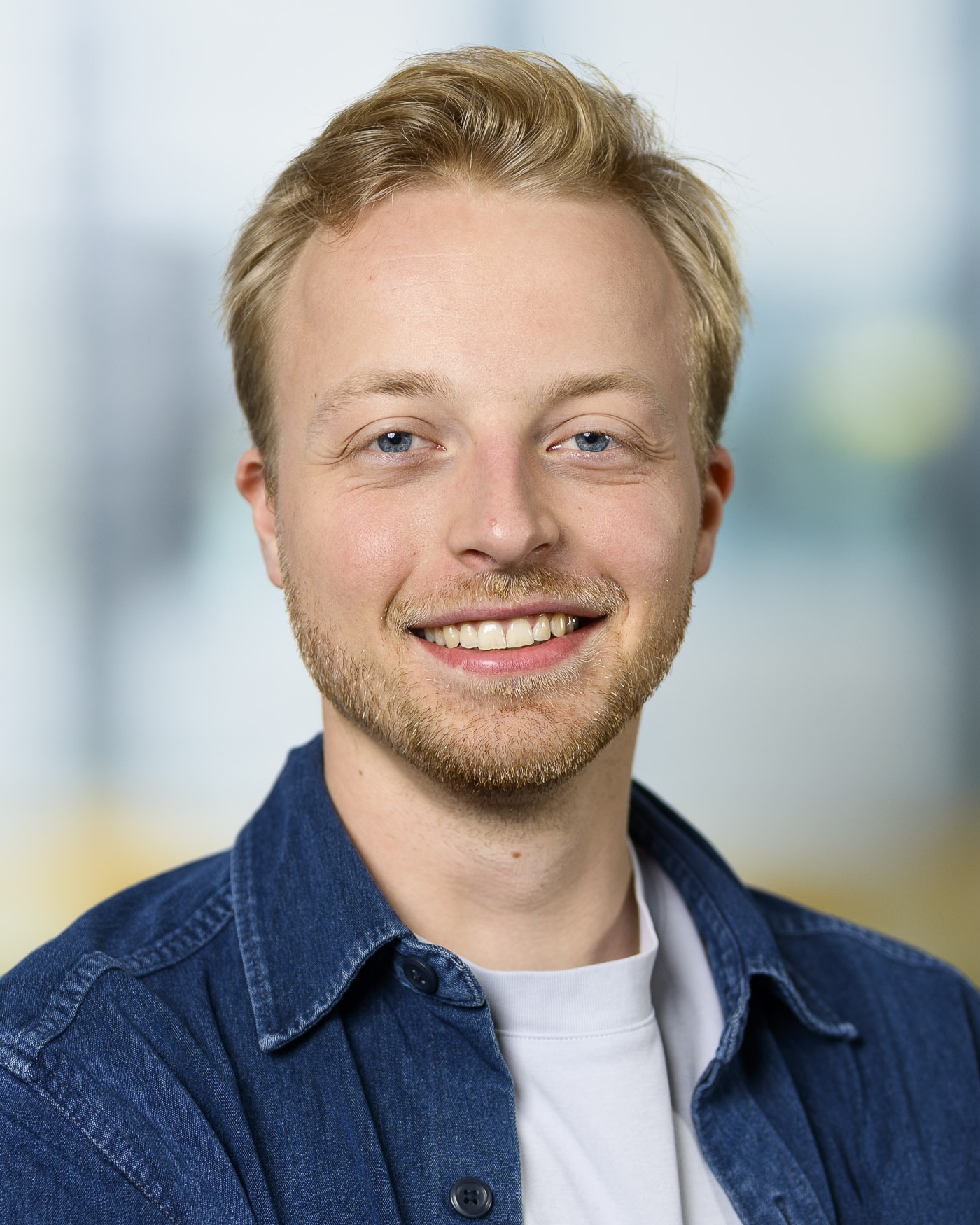}}]{Leon G.G. Olde Scholtenhuis}
received the M.Sc. degree in Applied Physics from Delft University of Technology (TU Delft). He is currently a Ph.D. candidate in the Terahertz Sensing Group at the same university, where he collaborates closely with the Space Research Organisation Netherlands (SRON) on device fabrication.

His research focuses on enhancing instrument performance through optimized nano- and microfabrication techniques. In this context, he is involved in the Terahertz Integrated Field Units with Universal Nanotechnology (TIFUUN) project, contributing to the development and fabrication of next-generation on-chip imaging spectrometers.
\end{IEEEbiography}

\begin{IEEEbiography}[{\includegraphics[width=1in,height=1.25in,
clip,keepaspectratio]{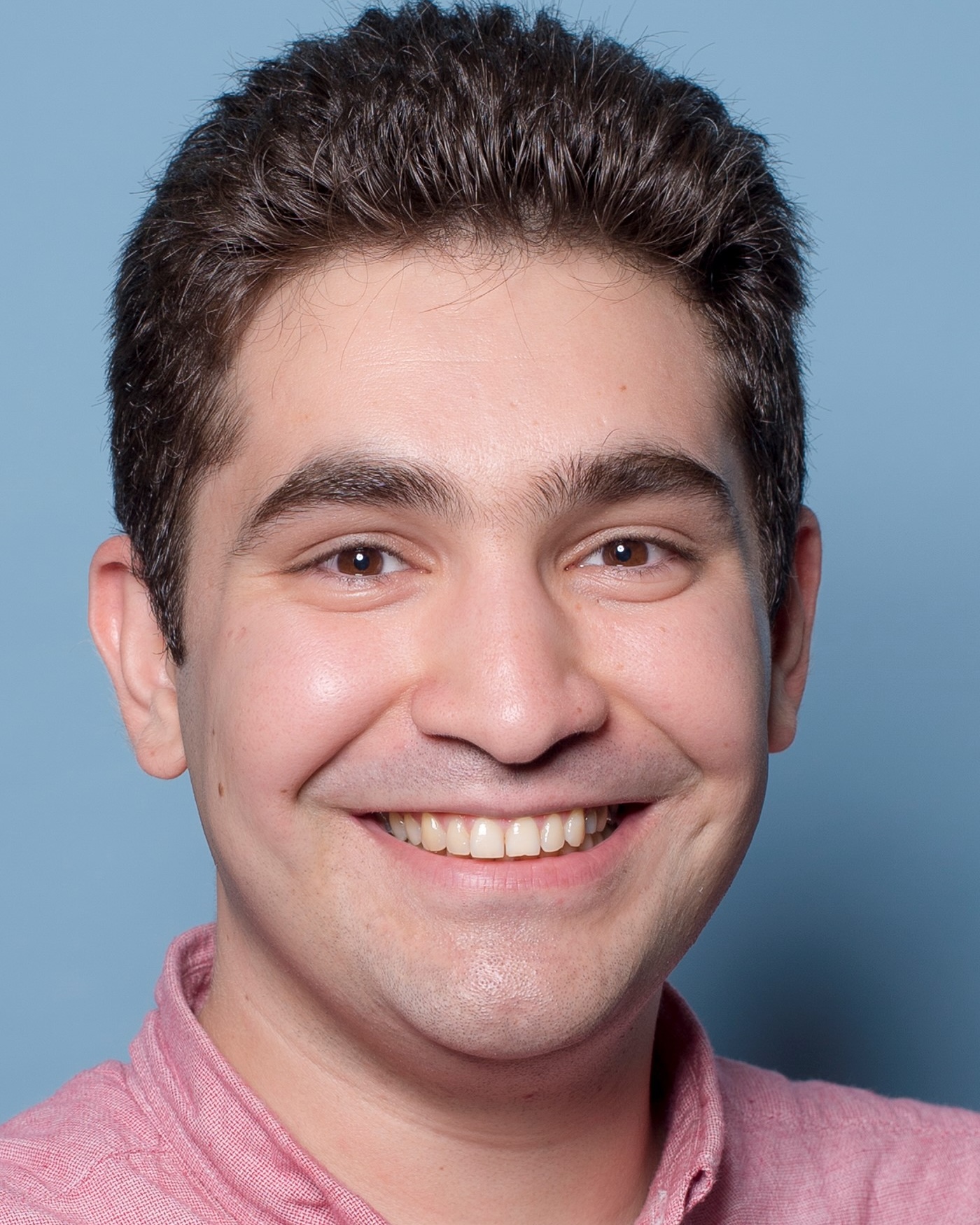}}]{Shahab O. Dabironezare} (Fellow, IEEE)
was born in Mashhad, Iran. He received his B.Sc. degree (cum laude) in Electrical Engineering-Communications from the Ferdowsi University of Mashhad (FUM), Mashhad, Iran, in 2013. He received the M.Sc. and Ph.D. degrees in electrical engineering from Delft University of Technology (TU Delft), Delft, The Netherlands, in 2015 and 2020, respectively.

From 2020 to 2022, he was a Post-Doctoral Researcher with the Department of Microelectronics, Terahertz (THz) Sensing Group, TU Delft. He is currently an Assistant Professor at the THz Sensing Group, TU Delft, and an Instrument Scientist at the Space Research Organisation Netherlands (SRON), Leiden, the Netherlands.
His research interests include wide-band antennas at millimetre and sub millimetre wave applications, lens absorbers and antennas for far-infra red astronomy, and wide field-of-view quasi-optical systems.
\end{IEEEbiography}

\begin{IEEEbiography}[{\includegraphics[width=1in,height=1.25in,
clip,keepaspectratio]{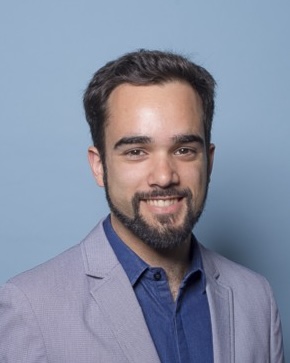}}]{Alejandro Pascual Laguna}
was born in Madrid, Spain, in 1992. He received the B.Sc. degree in telecommunications engineering from ICAI School of Engineering, Universidad Pontificia Comillas, Madrid, Spain, in 2014 after spending an exchange year at Chalmers University of Technology, Gothenburg, Sweden. He then obtained the M.Sc. (cum laude) and Ph.D. degrees in electrical engineering, respectively in 2016 and 2022, from the Delft University of Technology, Delft, The Netherlands.

From 2016 to 2023, he was with SRON, the Space Research Organization of the Netherlands, Leiden, The Netherlands; first as a Ph.D. candidate and then as a Scientist. Since 2023, he is with CAB, the Astrobiology Center (CSIC-INTA), Torrejón de Ardoz, Spain, where he is currently a Juan de la Cierva fellow. His research interests include on-chip solutions for efficient broadband (sub-)millimeter wavelength imaging spectrometers and polarimeters based on ultra-sensitive Kinetic Inductance Detectors.
\end{IEEEbiography}

\begin{IEEEbiography}[{\includegraphics[width=1in,height=1.25in,
clip,keepaspectratio]{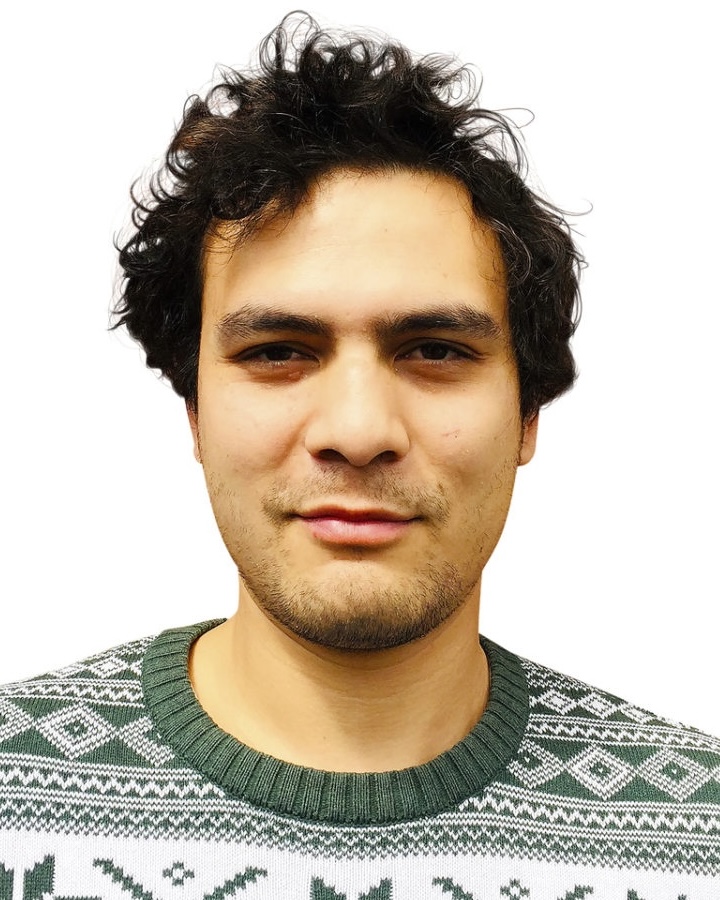}}]{Arend Moerman}
received the M.Sc. degree in astronomy and astrophysics from Leiden University, Leiden, The Netherlands, in 2022. He is currently with Delft University of Technology, where he is pursuing the Ph.D. degree.

His current research interests lie in commissioning novel instrumentation for the mm/sub-mm  wavelength range, galaxy cluster dynamics, and computational methods for astronomy.
\end{IEEEbiography}

\begin{IEEEbiography}[{\includegraphics[width=1in,height=1.25in,
clip,keepaspectratio]{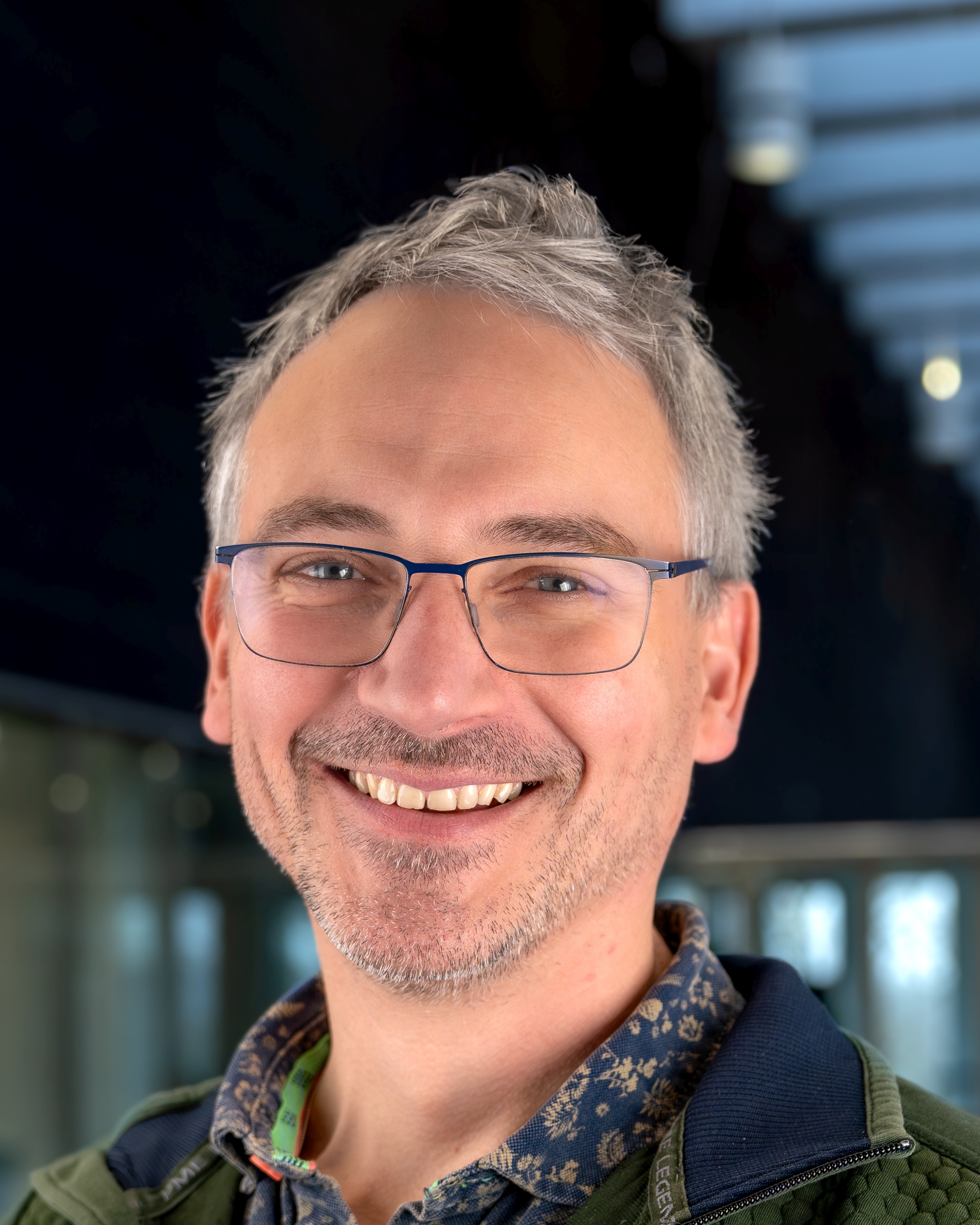}}]{David J. Thoen}
was born in the Netherlands in 1978. He received the BSc. degree in Applied Physics from the Fontys University of Applied Sciences Eindhoven, the Netherlands, in 2008, "met lof" (analogous to \textit{cum laude}). He conducted his research at FOM Institute for Plasma Physics Rijnhuizen, Nieuwegein, the Netherlands (now: DIFFER, Eindhoven).

From 2010 to 2015, he was a researcher engineer with the Kavli Institute of Nanoscience, Faculty of Applied Sciences, Delft University of Technology, The Netherlands. From 2015 till 2023, he has been with the Terahertz Sensing Group, Department of Microelectronics, Faculty of Electrical Engineering, Mathematics and Computer Science, Delft University of Technology. Since 2023 he is working in the Instrument Science Group, Space Research Organisation Netherlands, Leiden, The Netherlands. He has authored or coauthored 80 original articles published in international peer-reviewed journals, and various papers in conference proceedings.

Ing. Thoen is an experimental physisist working on the development and nanofabrication of superconducting electronics and optics and has expertise in deposition technologies of superconductors and dielectrics and high resolution patterning and imaging. He leads the Nano Development section at SRON.
\end{IEEEbiography}

\begin{IEEEbiographynophoto}{A. J. (Ton) van der Linden}
photograph and biography not available at the time of publication.
\end{IEEEbiographynophoto}

\begin{IEEEbiography}[{\includegraphics[width=1in,height=1.25in,
clip,keepaspectratio]{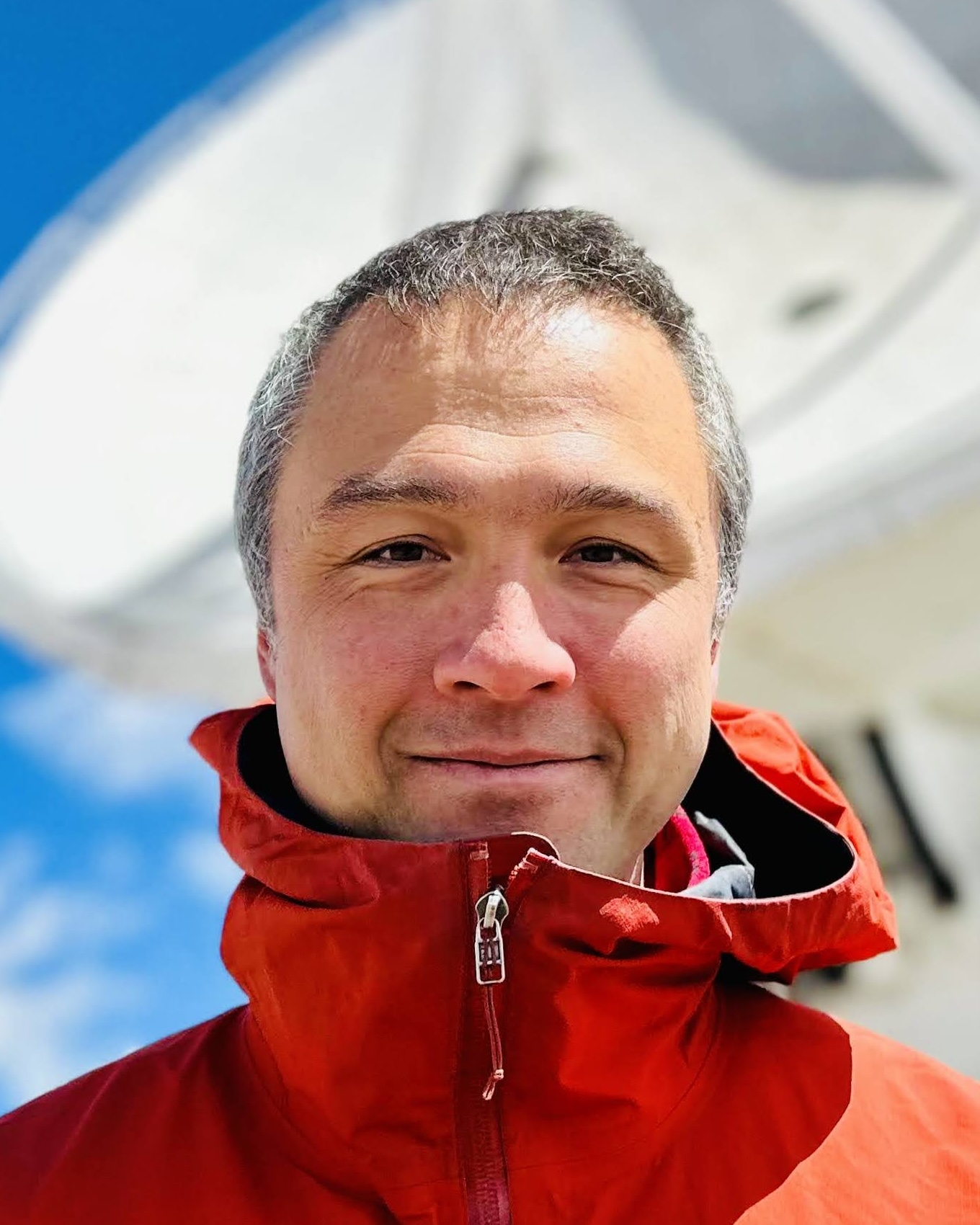}}]{Akira Endo}
was born in Japan in 1981. He received the Ph.D. degree from the School of Science, University of Tokyo, Japan, in 2009, with the Prize for Encouragement in Research (analogous to \textit{cum laude}). He conducted most of his doctoral research at the Advanced Technology Center, National Astronomical Observatory of Japan (NAOJ), Mitaka, Japan.

From 2009 to 2014, he was a Postdoctoral Researcher with the Kavli Institute of Nanoscience, Faculty of Applied Sciences, Delft University of Technology, The Netherlands. Since 2014, he has been with the Terahertz Sensing Group, Department of Microelectronics, Faculty of Electrical Engineering, Mathematics and Computer Science, Delft University of Technology, where he is currently an Associate Professor. He has authored or coauthored 58 original articles published in international peer-reviewed journals, two invited review articles, and numerous papers in conference proceedings.

Dr. Endo is an experimental astronomer whose work aims at mapping the large-scale structure of the Universe in three dimensions through the development of submillimeter-wave imaging spectrometers. He leads the DESHIMA collaboration as the Dutch Principal Investigator and currently leads a research program supported by the European Research Council (ERC) Consolidator Grant (2022) and the NWO Vici Grant (2026) to develop TIFUUN, an integral field spectrometer based on plug-and-play, open-hardware integral field units individually designed to address specific astronomical questions. He has previously received the NWO Veni, Vidi, NWO Medium Investment, and JSPS Overseas Research grants.
\end{IEEEbiography}

\begin{IEEEbiography}[{\includegraphics[width=1in,height=1.25in,
clip,keepaspectratio]{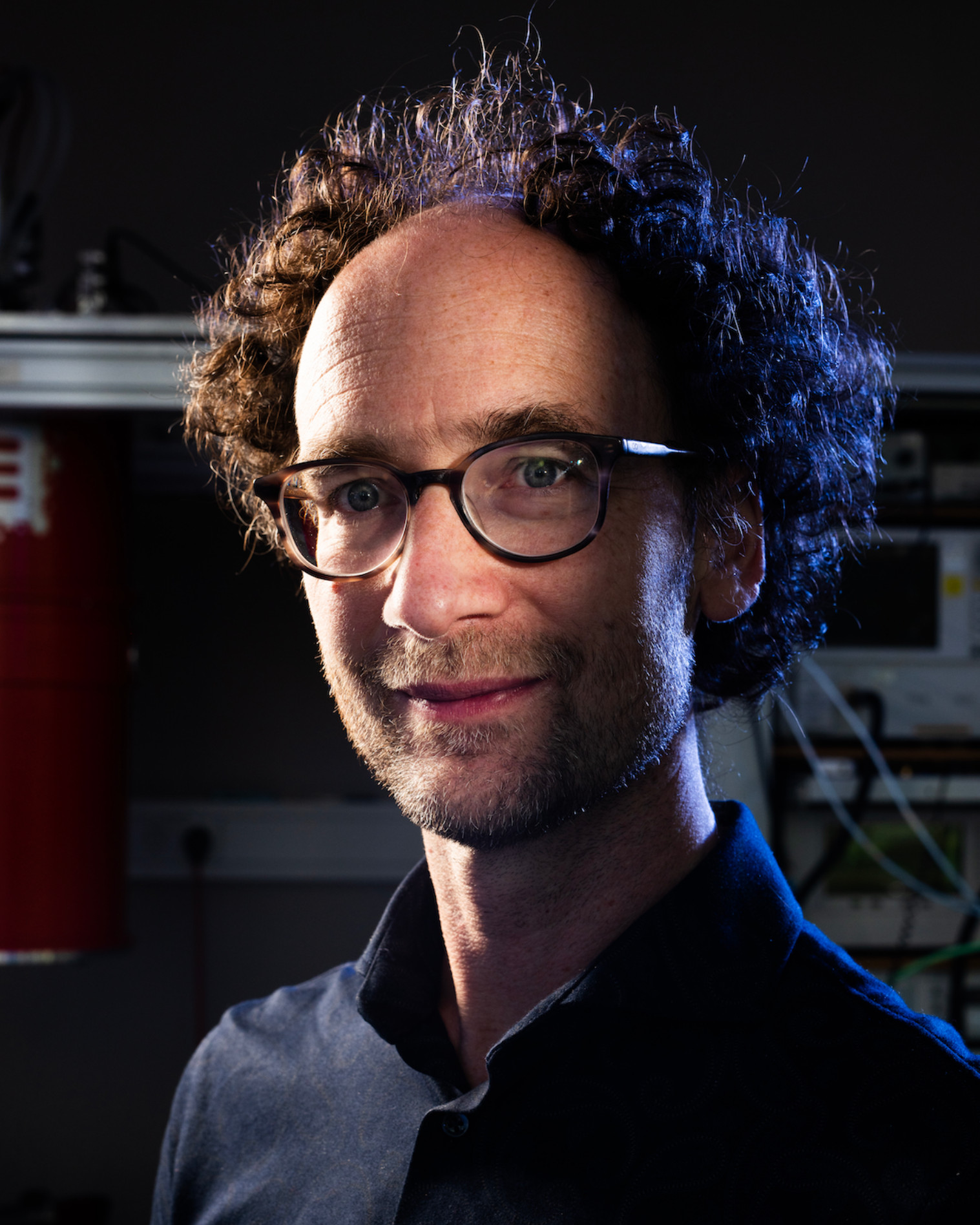}}]{Jochem J. A. Baselmans}
graduated in 1998 at the University of Groningen and received the Ph.D. degree at the university of Groningen in 2002 studying the superconducting state in normal metal Josephson junctions.

Prof. Dr. Ir. Jochem Baselmans is full professor in experimental astronomy at the Delft University of Technology, senior instrument scientist at the SRON Netherlands Institute for Space Research and member of the global faculty of the University of Cologne.

He is an expert in Kinetic Inductance Detectors (KIDs) and systems. He has pushed the sensitivity and TRL of KID systems to a level suitable for operation in cryogenically cooled space-based observatories. He is CO-I of PRIMA, as the lead for the detector development for its European instrument PRIMAger, detector lead for the POEMM balloon project and for the 18kpixel AMKID camera on the APEX telescope. Furthermore he pioneered the concept of the on-chip spectrometer and is the lead system engineer for the first on-chip spectrometer instruments, DESHIMA (2019) and DESHIMA-2 (2024).
\end{IEEEbiography}

\end{document}